\documentclass[prd,twocolumn,showpacs,showkeys,preprintnumbers,floatfix,
nofootinbib, superscriptaddress]{revtex4-1}
\usepackage{amsfonts} 
\usepackage{amssymb} 
\usepackage{amsmath} 
\usepackage{graphicx} 
\usepackage{subfigure} 
\usepackage{array} 
\usepackage{dcolumn} 
\usepackage{bm} 
\usepackage{latexsym} 
\usepackage{longtable} 
\usepackage{hyperref} 
\graphicspath{{./figs/}}
\renewcommand{\tilde}{\widetilde}

\newcommand{\dslash}[1]{\text{$\not \!\! #1$}}
\begin{document}


\title{Kaon mixing matrix elements from beyond-the-Standard-Model
operators in staggered chiral perturbation theory}
\author{Jon A. Bailey}
\affiliation{
 Lattice Gauge Theory Research Center, FPRD, and CTP \\
 Department of Physics and Astronomy,
 Seoul National University, 
 Seoul, 151-747, South Korea
}

\author{Hyung-Jin Kim}
%
%
\affiliation{
 Lattice Gauge Theory Research Center, FPRD, and CTP \\
 Department of Physics and Astronomy,
 Seoul National University, 
 Seoul, 151-747, South Korea
}
\author{Weonjong Lee}
\email[Email: ]{wlee@snu.ac.kr}
\homepage[Home page: ]{http://lgt.snu.ac.kr/}
%
%
\affiliation{
 Lattice Gauge Theory Research Center, FPRD, and CTP \\
 Department of Physics and Astronomy,
 Seoul National University, 
 Seoul, 151-747, South Korea
}
\author{Stephen R. Sharpe}
\email[Email: ]{sharpe@phys.washington.edu}
%
%
\affiliation{
 Physics Department, University of Washington, 
 Seattle, WA 98195-1560, USA \\
}
\collaboration{SWME Collaboration}
\date{\today}
\begin{abstract}
Models of new physics induce
$K-\overline{K}$ mixing operators having Dirac structures
other than the ``left-left'' form of the Standard Model.
We calculate the functional form of the corresponding $B$-parameters
at next-to-leading order in both SU(3) and SU(2)
staggered chiral perturbation theory (SChPT).
Numerical results for these matrix elements are being generated using improved
staggered fermions; our results can be used to extrapolate these matrix 
elements to the physical light and strange quark masses.
The SU(3) SChPT results turn out to be
much simpler than that for the Standard Model $B_K$ operator,
due to the absence of chiral suppression in the new operators.
The SU(2) SChPT result is of similar simplicity to that for $B_K$.
In fact, in the latter case,
the chiral logarithms for two of the new $B$-parameters
are identical to those for $B_K$, while those for
the other two new $B$-parameters are of opposite sign.
In addition to providing results for the $2+1$ flavor
theory in SU(3) SChPT and the $1+1+1$ flavor theory in
SU(2) SChPT, we present the corresponding continuum partially
quenched results, as these are not available in the literature.
\end{abstract}
\maketitle

\section{Introduction \label{sec:intr}}

Lattice calculations of the kaon mixing parameter $B_K$ 
are now very precise, with results using
several types of fermion showing reasonable consistency.
These results play an important role in constraining the
parameters of the Standard Model (SM)~\cite{Laiho:2009eu,Colangelo:2010et}.
Here we consider matrix elements of operators having
Dirac structures other than the ``left-left'' form that arises
in the SM.
These new operators generically appear in models of
physics beyond the SM (BSM) when heavy particles (e.g. squarks and gluinos
in supersymmetric models) are 
integrated out (see, e.g., 
Refs.~\cite{Ciuchini:1998ix,Buras:2001ra,Ciuchini:2006dw,Buras:2012fs}).
The resulting
$\Delta S=2$ four-fermion operators give rise to
additional contributions to both the CP-conserving and CP-violating
kaon mixing matrix elements, both of which are strongly constrained by
experimental results. Thus, if one can calculate the corresponding
hadronic matrix elements, one can place significant constraints
on BSM physics
(see, e.g., Refs.~\cite{Ciuchini:1998ix,Bona:2005eu,Isidori:2010kg}).

As part of the Staggered Weak Matrix Element (SWME) collaboration, we
are undertaking a numerical calculation
of these matrix elements using improved staggered fermions---specifically,
HYP-smeared~\cite{Hasenfratz:2001hp} 
valence quarks on lattice configurations
generated by the MILC collaboration with $N_f=2+1$ flavors of 
asqtad sea quarks~\cite{Bernard:2001av}.
To extrapolate or interpolate the results to the physical $d$ and $s$
quark masses and the continuum limit without introducing model
dependence, it is advantageous to 
use functional forms incorporating the constraints of chiral symmetry.
These forms can be calculated using staggered chiral perturbation theory 
(SChPT)~\cite{Lee:1999zxa,Bernard:2001yj,Aubin:2003mg}, 
and the present paper provides the results
at next-to-leading order (NLO).
One also needs to match the lattice operators onto those regularized
in a continuum scheme, and the necessary matching factors were
previously calculated to one-loop order in
perturbation theory~\cite{Kim:2011pz}.

The corresponding analysis for the $B_K$ operator has been
carried out in Ref.~\cite{VandeWater:2005uq}, and turns out to
be quite challenging. The difficulty arises because the
left-left structure of the four-fermion operator leads to
suppression of the $K-\overline{K}$
matrix element in the chiral limit. 
However, many of the operators which arise from mixing
due to discretization errors and truncated perturbative matching 
do not have chirally suppressed matrix elements.
This leads to a plethora of unknown low-energy constants (LECs)
entering at NLO in SU(3) SChPT. The situation is much simpler, however,
in SU(2) SChPT, where there is only one additional
LEC at NLO compared to continuum ChPT.

Similarly, the results for the BSM four-fermion operators
are much simpler than for the $B_K$ operator
because none of the BSM
operators have chirally suppressed matrix elements.
In fact, the situation for both SU(3) and SU(2) SChPT for the BSM operators
is the same as that for $B_K$ in SU(2) SChPT.
As long as one considers appropriate ratios (``$B$ factors''), 
there is only a single additional LEC
compared to the continuum ChPT expressions. This new LEC is induced
by matching and discretization errors. The simplicity of the
SChPT result should make the extrapolation of the new matrix elements
straightforward.

One-loop results in continuum ChPT for the BSM
operators have been given in Ref.~\cite{Becirevic:2004qd}.
These provide an important check on our results.
As a spin-off from our calculation, we provide the partially quenched
generalization of the results of Ref.~\cite{Becirevic:2004qd}
for both SU(3) and SU(2) ChPT.

This paper is organized as follows.
In Sec.~\ref{sec:OpsandBs} we list the BSM operators
and describe how they are mapped into the 
partially quenched (PQ) lattice theory.
In Sec.~\ref{sec:mapintoSChPT}, following
a brief review of SChPT, we explain the mapping
of the lattice operators into the chiral effective theory.
Section~\ref{sec:results} presents the NLO calculation
and results, first for SU(3) PQSChPT and then for SU(2) PQSChPT.
We also give the continuum SU(3) and SU(2) results in PQChPT.
We close in Sec.~\ref{sec:conc} with some brief conclusions.

\section{Operators and $B$-factors \label{sec:ops}}
\label{sec:OpsandBs}

\subsection{Continuum operators}

We use the so-called SUSY basis for the BSM
operators~\cite{Gabbiani:1996hi}
\begin{eqnarray}
O_2 &=& \bar{s}^a (1\!-\!\gamma_5) d^a \bar{s}^b (1\!-\!\gamma_5) d^b
\label{eq:O2cont}
,\\
O_3 &=& \bar{s}^a (1\!-\!\gamma_5) d^b \bar{s}^b (1\!-\!\gamma_5) d^a
\label{eq:O3cont}
,\\
O_4 &=& \bar{s}^a (1\!-\!\gamma_5) d^a \bar{s}^b (1\!+\!\gamma_5) d^b
\label{eq:O4cont}
,\\
O_5 &=& \bar{s}^a (1\!-\!\gamma_5) d^b \bar{s}^b (1\!+\!\gamma_5) d^a
\label{eq:O5cont}
,
\end{eqnarray}
where $a$ and $b$ are color indices.
Together with the $B_K$ operator
\begin{equation}
O_1 =  \bar{s}^a \gamma_\mu (1\!-\!\gamma_5) d^a 
\bar{s}^b \gamma_\mu (1\!-\!\gamma_5) d^b
\end{equation}
they form a complete set of $\Delta S=2$ four-fermion operators.
In the following we will concentrate entirely on the BSM operators,
and the index $j$ will always run over the values $2-5$.

These operators must be renormalized, and the Wilson coefficients which
multiply them are usually calculated in a canonical choice of
continuum scheme (such as $\overline{\rm MS}$ with naive dimensional
regularization of $\gamma_5$) at a canonical scale (such as $2\;$GeV).
Consequently, we are ultimately interested in the matrix elements of the operators
defined in such a scheme.  On the lattice, however, one inevitably starts with
bare lattice operators, or with operators defined in a regularization
independent scheme such as RI-MOM, and one must match the operators to those in
the canonical scheme. In our ongoing numerical work we presently use one-loop
perturbative matching to bare lattice operators, using the results from
Ref.~\cite{Kim:2011pz}.  Thus, in the following we assume that we are using
those linear combinations of lattice operators which match at one-loop order to
the operators defined in the canonical continuum scheme. 

It is convenient and conventional to package our ignorance
of the matrix elements of $O_{2-5}$ into $B$-parameters.
For the BSM operators these are defined as 
follows~\cite{Allton:1998sm}:
\begin{eqnarray}
 B_j (\mu) &=& 
\frac{\langle \overline{K}_0 | O_j(\mu) | K_0 \rangle}
     {N_j \langle \overline{K}_0 | \bar{s}^a \gamma_5 d^a(\mu)|0\rangle
      \langle 0 | \bar{s}^b \gamma_5 d^b(\mu)|K_0\rangle}
, \label{eq:Bj}
\end{eqnarray}
\begin{equation}
(N_2,N_3,N_4,N_5) = (5/3,-1/3,-2,-2/3)
,
\end{equation}
where $\mu$ is the renormalization scale.
The denominators in these ratios are obtained using the
vacuum saturation approximation, including the contribution
from the Fierz-rearranged form of the operators, but dropping
contributions suppressed in the chiral limit.
The product of matrix elements in the denominators can
be written as
\begin{equation}
\langle \overline{K}_0 | \bar{s} \gamma_5 d(\mu)|0\rangle
      \langle 0 | \bar{s} \gamma_5 d(\mu)|K_0\rangle
= - \left(\frac{f_K M_K^2}{m_d(\mu)+m_s(\mu)}\right)^2,
\label{eq:denom}
\end{equation}
which makes explicit that both the numerator and
denominator in the ratios defining the $B_j$ depend
on the renormalization scale, and should
be defined in a common scheme.

There are several advantages to using the ratios $B_j$
rather than directly calculating the matrix elements
$\langle \overline{K}_0 | O_j(\mu) | K_0 \rangle$.
First, in a lattice calculation, forming a dimensionless
ratio reduces statistical and systematic errors---particularly
those due to the uncertainty in the lattice spacing and in
the matching factors. Second, the SWME lattice calculation
uses wall sources, following the same methodology as
for $B_K$~\cite{Bae:2010ki}, and 
the overlap factor between these sources and the kaon states
cancels in the ratio. Third, as we will see below, the
SChPT expression for the ratio is simpler (involving fewer LECs)
than for the matrix elements.
There is, however, also a potential disadvantage,
as stressed in Ref.~\cite{Donini:1999nn}.
To convert from the
$B_j$ to the corresponding matrix elements, one must
multiply by the denominator, which, as shown by
Eq.~(\ref{eq:denom}), depends on $m_d$ and $m_s$.
These quark masses are not directly measurable physical quantities, and
must be obtained from lattice calculations.
In the late 1990's, when Ref.~\cite{Donini:1999nn} was written,
there were large systematic errors in determinations of
light quark masses; the source of largest error was the
quenched approximation. The present situation
is markedly improved, with quark masses known to about
1\% accuracy~\cite{Laiho:2009eu,Colangelo:2010et}.
Thus, there is no longer a phenomenological reason
not to use the $B_j$.

\subsection{Lattice operators in the continuum limit}
\label{sec:Olat_cont}

To calculate the $B_j$ with staggered fermions, we must
account for the additional taste degree of freedom.
Each quark flavor enlarges to a quartet with four tastes.
In this subsection, we first consider the
staggered theory in the continuum limit, in which the
taste symmetry is exact.\footnote{%
We are assuming here that the rooting procedure used to
remove the additional tastes from the quark-gluon sea defines
a theory with the correct continuum limit.}
Taste-breaking corrections will be discussed in the next
subsection.

The additional tastes occur both in the operators $O_j$ and
in the external kaons. We choose the latter to have taste $P$,
i.e. to be created by operators with spin-taste $\gamma_5\otimes \xi_5$. 
This kaon is the pseudo-Goldstone boson (PGB) associated
with the spontaneous breaking of an axial $U(1)_A$ symmetry 
which holds exactly on the lattice in the massless limit. 
It follows that its correlation functions satisfy Ward-Takahashi identities
which are analogues of those in the continuum~\cite{Kilcup:1986dg}.
This in turn leads to simplifications in the SChPT expressions 
for its matrix elements.  The taste $P$ kaon is also the simplest choice for
numerical calculations, since it is the lightest kaon state.

Turning now to the operators, we face the problem that Fierz
transformations of the continuum $O_j$ are no longer 
matched by those of the lattice operators once we introduce
taste. Following Refs.~\cite{Kilcup:1997ye,VandeWater:2005uq}, 
we resolve this
problem by introducing two types each of valence $s$ and $d$ quarks.
We label these $S_1$ and $S_2$ (or $D_1$ and $D_2$) using uppercase
letters to denote fields which include the taste degree of freedom.
Then the operators in the continuum staggered theory take the
form~\cite{Kim:2011pz}
\begin{align}
 \mathcal{O}_{2}^{{\rm Cont}'}&=
 \mathcal{O}^{{\rm Cont}'}_{S,II}\!\!\!+\!
 \mathcal{O}^{{\rm Cont}'}_{P,II}
\!\!-\! \frac12 \left(\mathcal{O}^{{\rm Cont}'}_{S,I}\!\!+\!
 \mathcal{O}^{{\rm Cont}'}_{P,I}\!\!\!-\!
 \mathcal{O}^{{\rm Cont}'}_{T,I}\right),
\label{eq:O2cont'}\\
 \mathcal{O}_{3}^{{\rm Cont}'}&=
 \mathcal{O}^{{\rm Cont}'}_{S,I}\!\!+\!
 \mathcal{O}^{{\rm Cont}'}_{P,I}
\!\! -\! \frac12 \left(\mathcal{O}^{{\rm Cont}'}_{S,II}\!\!+\!
 \mathcal{O}^{{\rm Cont}'}_{P,II}\!\!-\!
 \mathcal{O}^{{\rm Cont}'}_{T,II}\right),
\label{eq:O3cont'}\\
 \mathcal{O}_{4}^{{\rm Cont}'}&=
 \mathcal{O}^{{\rm Cont}'}_{S,II}\!\!\!-\!
 \mathcal{O}^{{\rm Cont}'}_{P,II}
\!\! -\! \frac12 \left(\mathcal{O}^{{\rm Cont}'}_{V,I}\!\!\!-\!
 \mathcal{O}^{{\rm Cont}'}_{A,I}\right),
\label{eq:O4cont'}\\
 \mathcal{O}_{5}^{{\rm Cont}'}&=
 \mathcal{O}^{{\rm Cont}'}_{S,I}\!\!\!-\!
 \mathcal{O}^{{\rm Cont}'}_{P,I}
\!\! -\! \frac12 \left(\mathcal{O}^{{\rm Cont}'}_{V,II}\!\!\!-\!
 \mathcal{O}^{{\rm Cont}'}_{A,II}\right).
\label{eq:O5cont'}
\end{align}
Here the subscripts indicate firstly the ``spin'' of the
four-fermion operator and secondly the manner in which
the color indices are contracted.\footnote{%
The notation is slightly changed from that in Ref.~\cite{Kim:2011pz}
so as to conform to the more convenient 
notation of Ref.~\cite{VandeWater:2005uq}.}
The prime in the superscript ${\rm Cont}'$ 
is a reminder that this is the continuum theory in which
the number of valence $s$ and $d$ quarks have been doubled.

The ``two-color-loop'' operators (denoted by subscripts ``$II$'') are
\begin{align}
 \mathcal{O}^{{\rm Cont}'}_{S,II} &\equiv
\bar S_1^a ({\bf 1} \otimes \xi_5) D_1^a\;
\bar S_2^b({\bf 1} \otimes \xi_5) D_2^b , 
\label{eq:OSII}
\\
 \mathcal{O}^{{\rm Cont}'}_{P,II} &\equiv
\bar S_1^a (\gamma_5 \otimes \xi_5) D_1^a\;
\bar S_2^b(\gamma_5 \otimes \xi_5) D_2^b , 
\\
 \mathcal{O}^{{\rm Cont}'}_{T,II} &\equiv
\sum_{\mu<\nu} \bar S_1^a (\gamma_\mu\gamma_\nu \otimes \xi_5) D_1^a\;
\bar S_2^b(\gamma_\mu\gamma_\nu \otimes \xi_5) D_2^b ,
\label{eq:OTII}
\\
 \mathcal{O}^{{\rm Cont}'}_{V,II} &\equiv
\sum_\mu \bar S_1^a (\gamma_\mu \otimes \xi_5) D_1^a\;
\bar S_2^b(\gamma_\mu\otimes \xi_5) D_2^b ,
\\
 \mathcal{O}^{{\rm Cont}'}_{A,II} &\equiv
\sum_{\mu} \bar S_1^a (\gamma_\mu\gamma_5 \otimes \xi_5) D_1^a\;
\bar S_2^b(\gamma_\mu\gamma_5 \otimes \xi_5) D_2^b ,
\end{align}
and are so named because, when contracted with external color-singlet
kaon fields, there are two loops of color indices.
The corresponding ``one-color-loop'' operators differ only
in their color indices, as exemplified by
\begin{equation}
 \mathcal{O}^{{\rm Cont}'}_{S,I} \equiv
\bar S_1^a ({\bf 1} \otimes \xi_5) D_1^b\;
\bar S_2^b({\bf 1} \otimes \xi_5) D_2^a .
\label{eq:OSI}
\end{equation}

The matrix elements of the operators ${\cal O}_j^{{\rm Cont}'}$
are to be taken between a taste-$P$ kaon of type 2, $K_{P2}^{0}$,
created by the operator $\bar D_2(\gamma_5\otimes \xi_5) S_2$, and an
antikaon of type 1, $\overline{K}_{P1}^{0}$, 
destroyed by $\bar D_1(\gamma_5\otimes \xi_5) S_1$.
In this way we force the component operators of the ${\cal O}_j^{{\rm Cont}'}$
to have a single Wick contraction with the external kaon fields,
and thus avoid the Fierz-transformed contractions which occur
in the matrix elements of the original operators $O_j$.
The latter contractions are then added back by hand, giving rise
to the terms in parentheses in 
Eqs.~(\ref{eq:O2cont'}-\ref{eq:O5cont'}).
Note that since the external kaons have taste $P$, the
bilinears composing the four-fermion operators
in Eqs.~(\ref{eq:OSII}-\ref{eq:OSI}) must also have
this taste, since taste is a good symmetry in the staggered
continuum theory. 

The correspondence between $B$-parameters in QCD and those
in the augmented staggered theory can now be given.
In the continuum limit of the latter theory, we have
\begin{equation}
B_j 
= \frac{2 \langle\overline{K}_{P1}^{0}|{\cal O}_j^{{\rm Cont}'}|K_{P2}^{0}\rangle}
{N_j \langle\overline{K}_{P1}^{0}|\bar S_1(\gamma_5\otimes \xi_5) D_1|0\rangle
     \langle 0|\bar S_2(\gamma_5\otimes \xi_5) D_2 |K_{P2}^{0}\rangle}
.
\label{eq:Bjlatt}
\end{equation}
In essence, we have constructed lattice operators which have the same
Wick contractions with the external fields as do the original operators
$O_j$ between physical kaons. The extra factor of $2$ in the
numerator [compared to Eq.~(\ref{eq:Bj})] accounts for the
fact that in the original theory each bilinear could be contracted
with either external field, whereas here there is only one such
contraction due to the presence of two types of
$S$ and $D$ quarks.
In the staggered theory one also must account for possible
factors of the number of tastes, $N_t=4$. 
Such factors cancel in the ratios $B_j$~\cite{VandeWater:2005uq}, 
and so are not shown explicitly.

At this stage, it is helpful to summarize the content of
the augmented staggered theory that we have constructed.
This theory contains $U$, $D$ and $S$ sea quarks,
as well as $D_1$, $D_2$, $S_1$ and $S_2$ valence quarks.
Each of these fields represents four degenerate tastes.
We allow the masses of the sea and valence quarks to differ,
since we make use of this freedom in our simulations. 
We call the sea quark masses $m_u$, $m_d$ and $m_s$,
respectively, while we follow Ref.~\cite{VandeWater:2005uq} and
denote $m_{d1}=m_{d2}=m_x$ and $m_{s1}=m_{s2}=m_y$.
Note that we choose both strange valence quarks to have the
same mass, and similarly for the down valence quarks.
Finally, we must add ghost quarks for each of the valence quarks,
and apply the fourth-root prescription to the sea-quark determinant.
In the continuum limit, rooting is equivalent to adding 3 tastes
of ghost quark for each sea-quark field.
Including both flavor and taste in the counting, the
resulting partially quenched (PQ) theory has 28 quarks and 25 ghost-quarks. 
This is the minimal field content required to represent the
desired operators when using rooted staggered fermions.

Although the construction of this PQ theory has been motivated
by our use of staggered lattice fermions, one can also
consider it as a purely continuum theory with no reference
to the lattice. The result (\ref{eq:Bjlatt}) still holds,
and is a relationship between matrix elements in two different
continuum theories, one unquenched and the other partially quenched.
If one regularizes these two theories in the same way, then
the relationship holds for all values of the renormalization scale $\mu$.
In particular, the anomalous dimension matrix of the four operators
${\cal O}_j^{{\rm Cont}'}$ should be the same as that of the
original operators $O_j$. The results of Ref.~\cite{Kim:2011pz}
check this explicitly at one-loop order.\footnote{%
Strictly speaking, one must use a regularization which preserves
the Fierz identities, such as the RI-MOM 
scheme of Refs.~\cite{Ciuchini:1997bw,Buras:2000if}
or the $\overline{\rm MS}$ scheme proposed in Ref.~\cite{Gupta:1996yt}. 
}

It will be useful in the following to consider also matrix
elements of ${\cal O}_j^{{\rm Cont}'}$ between kaons having
different flavors and tastes. First we note that,
because taste is a good symmetry in the continuum limit of
the lattice theory, the matrix elements between type 1 and 2
kaons vanish unless they have taste $P$:
\begin{equation}
\langle\overline{K}_{B1}^{0}|{\cal O}_j^{{\rm Cont}'}|K_{B2}^{0}\rangle
= 0 \quad {\rm if} \quad B\ne P
.
\label{eq:tasteI}
\end{equation}
Here $B$ labels one of the 16 choices of taste for the
external kaons, as will be described shortly.
Second, we consider matrix elements between mixed flavor
kaons. Let $K^0_{B12}$ be the kaon created by
$\bar D_1(\gamma_5\otimes\xi_B) S_2$, and
$\overline{K}^0_{B21}$ be the antikaon destroyed by
$\bar D_2(\gamma_5\otimes\xi_B) S_1$.
Here we are labeling tastes by a hypercube vector
$B=(B_1,B_2,B_3,B_4)$, in which each entry is either $0$ or $1$,
and 
\begin{equation}
\xi_B = \xi_1^{B_1} \xi_2^{B_2} \xi_3^{B_3} \xi_4^{B_4},
\qquad (\xi_\mu=\gamma_\mu^*).
\end{equation}
Thus, for example, $B=(1,1,1,1)$ corresponds to taste $P$.
We then find that, for each value of $j$,
\begin{equation}
\langle\overline{K}_{B21}^{0}|{\cal O}_j^{{\rm Cont}'}|K_{B12}^{0}\rangle
= \frac{s_B}{4}
\langle\overline{K}_{P1}^{0}|{\cal O}_j^{{\rm Cont}'}|K_{P2}^{0}\rangle
,
\label{eq:FierzO}
\end{equation}
where the sign $s_B$ is 
\begin{equation}
s_B = \frac14 {\rm tr}\left(
\xi_B \xi_5 \xi_B \xi_5\right)
.
\end{equation}
This result is obtained by Fierz transforming the
operators in order to bring the bilinears into
an ``$(\bar S_1 D_2)(\bar S_2 D_1)$'' form.
One must simultaneously Fierz-transform in color,
spin and taste. While the operators in 
Eqs.~(\ref{eq:O2cont'}-\ref{eq:O5cont'}) are,
by construction, Fierz-invariant in color and spin, they are
not Fierz-invariant in taste. Taste $P$ Fierz-transforms
into all tastes, with the weight factor being $s_B/4$.
We stress again that the result (\ref{eq:FierzO})
holds only in the continuum limit, for it relies
on having an exact taste symmetry.

\subsection{Lattice operators for $a\ne 0$}
\label{subsec:Olat_for_a}

Numerical calculations of the matrix elements
required for Eq.~(\ref{eq:Bjlatt}) are being carried out
in a lattice theory with three rooted sea quarks
and two flavors each of valence down and strange quarks.
This theory provides a lattice regularization of
the PQ continuum theory described in the previous subsection.
In this subsection we discuss the impact of the
discretization errors inherent in the lattice regularization
on the extraction of the desired matrix elements.
The dominant issue is the presence of taste-symmetry
breaking for $a\ne0$.

To start with, we must choose a discretization of
the continuum operators. The simplest choice is to
use operators living on a $2^4$ hypercube, using
the method of Ref.~\cite{KlubergStern:1983dg} to obtain operators with
the desired spins and taste. 
We call the resulting operators ${\cal O}_j^{\rm Lat}$.
The details of our particular
implementation have been described in Ref.~\cite{Kim:2011pz}
and will not be important.
What matters here is the structure of the
matching between lattice operators and
those defined in the PQ continuum theory.
The general form for the four-fermion operators is\footnote{%
There are no contributions proportional to $a$ because these would
arise from mixing with dimension 7 operators, but, as explained
in Ref.~\cite{VandeWater:2005uq}, such operators have the
wrong tastes to contribute to the desired matrix elements.}
\begin{eqnarray}
{\cal O}_j^{{\rm Cont}'} \!\!\!\!&\cong&\!\! {\cal O}_j^{\rm Lat}
+ \frac{\alpha}{4\pi} [\textrm{taste $P$ ops.}]
+ \frac{\alpha}{4\pi} [\textrm{other taste ops.}]
\nonumber \\ &&\!\!\!\!
+ \alpha^2 [\textrm{various taste ops.}] 
+ a^2 [\textrm{various taste ops.}]
\nonumber \\ &&
+ \dots
,
\label{eq:Ojmatch}
\end{eqnarray}
where the ellipsis indicates terms of higher order in $\alpha$ and $a$.
The symbol $\cong$ means here the equality of the matrix elements of the
operators on both sides of this equation, evaluated
in their respective theories (PQ continuum on the left-hand side,
lattice theory on the right).
Thus all operators on the right-hand side are
{\em lattice} four-fermion operators, and ``taste $B$''
indicates that {\em both} bilinears in the operator have this taste.
The expression ``various taste ops.'' 
implies that there are operators both with taste $P$
and with other tastes. The set of operators which
can appear is determined by the lattice symmetry group.

We have separated out the one-loop contributions in (\ref{eq:Ojmatch})
because they have been calculated in Ref.~\cite{Kim:2011pz},
matching to a Fierz-invariant $\overline{\rm MS}$ scheme in
the continuum PQ theory.
We use $\alpha/(4\pi)$ (rather than just $\alpha$) because the
largest one-loop coefficients are of ${\cal O}(1)\times \alpha/(4\pi)$.
For the two-loop terms we do not know {\em a priori} whether
the corrections are $\sim \alpha^2$ or $\sim [\alpha/(4\pi)]^2$
so we make the conservative choice and assume the former.
Given the numerical value of $\alpha(1/a)$ 
for present lattice spacings
(in, say, the $\overline{\rm MS}$ scheme), 
it is argued in Ref.~\cite{VandeWater:2005uq} that an
appropriate phenomenological
power counting is 
$\alpha/(4\pi)\sim \alpha^2 \sim (a\Lambda_{\rm QCD})^2\ll 1$.
We adopt this power counting here,
so all the displayed correction terms in Eq.~(\ref{eq:Ojmatch})
are formally of the same (small) size.

We have further separated in (\ref{eq:Ojmatch}) the 
one-loop contributions from operators with taste $P$
from those from operators with other tastes.
This is because, in our companion numerical calculations,
we explicitly include (for practical reasons) 
only the taste $P$ one-loop contributions.
In other words, the actual lattice four-fermion operator we
use is
\begin{equation}
{\cal O}_j^{{\rm Lat,Actual}}
 =
{\cal O}_j^{{\rm Lat}} 
+\frac{\alpha}{4\pi} [\textrm{taste $P$ ops.}]
.
\end{equation}
Moving the other contributions in Eq.~(\ref{eq:Ojmatch})
from the lattice to the continuum side of the equation
(which can be done using tree-level matching as these
contributions are of NLO due to the explicit factors of
$\alpha/4\pi$, $\alpha^2$ and $a^2$) we end up with 
\begin{eqnarray}
{\cal O}_j^{{\rm Lat,Actual}}
&\cong&\!\!
{\cal O}_j^{{\rm Cont}'} 
-\frac{\alpha}{4\pi} [\textrm{other taste ops.}]
\nonumber\\
&&\!\!
-\alpha^2 [\textrm{various taste ops.}] 
\nonumber\\
&&\!\!
-a^2 [\textrm{various taste ops.}] + \dots
\label{eq:Ojlaterror}
\end{eqnarray}
Here operators to the right of the $\cong$ are
now {\em continuum} four-fermion operators.
We see that our lattice operator corresponds in
the PQ continuum theory to the operator we want
together with several undesired operators.

A similar analysis can be done for the bilinear operators
appearing in the denominator of Eq.~(\ref{eq:Bjlatt}).
This case is simpler because, 
to all orders in perturbation theory,
there is no mixing with other bilinears, 
due to the  lattice symmetries~\cite{Verstegen:1985kt}.
Again, in practice we use a one-loop corrected operator,
which can be written (for $k=1,2$)
\begin{eqnarray}
\lefteqn{\left[\bar S_k(\gamma_5\otimes \xi_5) D_k\right]^{\rm Lat,Actual}
\cong 
\left[\bar S_k(\gamma_5\otimes \xi_5) D_k\right]^{{\rm Cont}'}}
\nonumber\\
\!\!\!\!\!&-&\!\!
\alpha^2 c \left[\bar S_k(\gamma_5\otimes \xi_5) D_k\right]
\!-\!a^2 [\textrm{various taste ops.}]\,
\label{eq:Platerror}
\end{eqnarray}
with $c$ an unknown constant of ${\cal O}(1)$.
In this case there are no errors proportional to $\alpha/(4\pi)$.

It is straightforward, although tedious, to
enumerate the operators which appear in 
Eqs.~(\ref{eq:Ojlaterror}) and (\ref{eq:Platerror}) in the terms
proportional to $\alpha/(4\pi)$, $\alpha^2$ and $a^2$.
For the $\alpha/(4\pi)$ terms, the full list
has been given in Ref.~\cite{Lee:2003sk},
along with their one-loop coefficients.\footnote{%
The coefficients are given in Ref.~\cite{Lee:2003sk}
only for the Wilson gauge action, rather than for the
improved Symanzik gauge action used in practice.
The results differ little, however~\cite{KLSprivate}.
In particular, the same operators have the
largest coefficients in both cases.}
%
For the other operators, one must use lattice symmetries,
and appropriately generalize the analysis given for
the $B_K$ operator in Ref.~\cite{VandeWater:2005uq}.
This exercise turns out, however, to be unnecessary when 
considering the $B_j$ at NLO in SChPT.
To explain this conclusion we must turn to the issue of
mapping operators into the chiral effective theory.

\section{Mapping operators into SChPT}
\label{sec:mapintoSChPT}

\subsection{Review of SChPT}
\label{subsec:SChPT}

We begin with a brief review of the relevant aspects
of SChPT. More details are given in Refs.~\cite{Aubin:2003mg}
and \cite{Sharpe:2004is}.
It is an effective theory constructed in three steps.
First, one determines
the Symanzik continuum effective Lagrangian describing
the interactions of quarks and gluons with $p\ll 1/a$,
which incorporates the leading discretization errors proportional
to $a^2$. 
Second, one maps the resulting theory into its chiral
counterpart, in which the degrees of freedom are the
pseudo-Goldstone particles produced by spontaneous chiral symmetry
breaking. It is straightforward to do this mapping only for
an unrooted theory, i.e. one in which one keeps all tastes as
dynamical degrees of freedom.
The final stage is to account for the rooting of the quark
determinant by including
appropriate factors of $1/4$ by hand for diagrams containing sea-quark
loops. This last stage has been put
on a firm theoretical footing by the work of 
Refs.~\cite{Bernard:2006zw,Bernard:2007ma}.

The standard power counting in SChPT is
$p^2\sim m\sim a_\alpha^2$.
Here $a_\alpha^2\equiv a^2 \alpha_V(\pi/a)^2$ is the size
of the leading taste-breaking corrections with HYP, asqtad
or HISQ fermions. As described above,
when one considers matrix elements one must
also include taste-conserving discretization errors proportional
to $a^2$ (without factors of $\alpha$ since HYP fermions 
are not fully improved) and matching errors proportional to $\alpha/(4\pi)$ and
$\alpha^2$. In the extended power counting introduced 
in Ref.~\cite{VandeWater:2005uq} one assumes
\begin{equation}
p^2\sim m\sim a_\alpha^2\sim a^2\sim \frac{\alpha}{4\pi}\sim \alpha^2
.
\end{equation}
We stress that the peculiar-looking choices
$a_\alpha^2\sim a^2$ and $\alpha/(4\pi)\sim\alpha^2$ are particular
to the case at hand and are phenomenologically based.
The choice $a_\alpha^2\sim a^2$ is made because it is found that 
taste-breaking discretization errors are numerically enhanced,
and only after suppression by $\alpha^2$ are they comparable to
other discretization errors.
As explained in Sec.~\ref{subsec:Olat_for_a}, the choice
$\alpha/(4\pi)\sim\alpha^2$ is based on the explicit results for one-loop
matching coefficients. 

The Symanzik continuum theory obtained in the first of the
steps described above is a partially quenched theory
containing 28 quarks (3 sea and 4 valence, each with 4 tastes)
and 16 ghost quarks. It is convenient to collect the corresponding
fields into a column-vector $Q$. In the combined chiral and continuum
limit, the Symanzik action has a graded chiral symmetry,
$SU(28|16)_L\times SU(28|16)_R$. To display this we define
left and right-handed Euclidean fields as usual, e.g.
$Q_L=(1\!-\!\gamma_5)/2 Q$ and $\bar Q_R = \bar Q (1\! -\!\gamma_5)/2$,
so that
\begin{equation}
{\cal L}_{\rm Sym} \stackrel{m,a\to 0}{\longrightarrow}
\bar Q_R \dslash{D} Q_R + \bar Q_L \dslash{D} Q_L.
\end{equation}
The symmetry is 
\begin{equation}
Q_L \to L Q_L,\ Q_R \to R Q_R,\ 
\bar Q_L \to \bar Q_L L^\dagger,\ 
\bar Q_R \to \bar Q_R R^\dagger,
\label{eq:Qtransform}
\end{equation}
with $L,R\in SU(28|16)_{L,R}$.
This graded symmetry is spontaneously broken down to
its diagonal subgroup, leading to $44^2-1$ pseudo-Goldstone
particles.\footnote{%
The grading does lead to some subtleties in the analysis of symmetries
and their implications, but
these do not effect perturbative calculations in the resulting
chiral theory~\cite{Sharpe:2000bc}.}

The chiral effective theory contains only the light
Goldstone particles that result after symmetry breaking.
These are collected as usual into
a $U(28|16)$ matrix $\Sigma=\exp(i\Phi/f)$
(with $f$ such that $f_\pi \approx 132\;$MeV), where
\begin{equation}
\Phi = \left(
\begin{array}{cccc}
    U & \pi^+ & K^+ & \cdots \\
\pi^- & D     & K^0 & \cdots \\
K^-   &\overline{K}^0 & S & \cdots\\
\vdots&\vdots &\vdots&\ddots
\end{array}
\right)
.
\end{equation}
Here each entry in the matrix is a $4\times4$ block
corresponding to the 16 different tastes.
Under the chiral symmetry, $\Sigma$ transforms as
\begin{equation}
\Sigma \longrightarrow L\Sigma R^\dagger.
\label{eq:Sigmatransform}
\end{equation}
The LO chiral Lagrangian is
\begin{eqnarray}
{\cal L}_\chi &=& \frac{f^2}8 {\rm str}\left(
\partial_\mu\Sigma\partial_\mu\Sigma^\dagger\right)
- \frac{B_0 f^2}{4} {\rm str}\left(
{\cal M}\Sigma + {\cal M}\Sigma^\dagger\right)
\nonumber\\
&&+ \frac{m_0^2}{24} [{\rm str}(\Phi)]^2 + a^2({\cal U}+ {\cal U'}) ,
\end{eqnarray}
where ``str'' stands for supertrace or ``strace'',
and ${\cal M}$ is the mass matrix
\begin{equation}
{\cal M} = {\rm diag}(m_u,m_u,m_u,m_u,m_d,m_d,m_d,m_d,m_s,\cdots)
.
\end{equation}
The $m_0$ term represents the effect of the axial anomaly
(with normalization as in Ref.~\cite{Aubin:2003mg});
$m_0$ is to be sent to infinity to remove the unwanted non-Goldstone
singlet particle~\cite{Sharpe:2001fh}.
The last term is the taste-breaking
potential arising from discretization errors.\footnote{%
Although this contribution is proportional to $a_\alpha^2$,
it is conventional to pull out an overall factor of just $a^2$.
This is a purely notational matter, since the difference can
be absorbed in the LECs contained in ${\cal U}$ and ${\cal U}'$.}

The taste-breaking potential consists of a single strace
component ${\cal U}$ and a double strace part ${\cal U'}$.
The former is
\begin{eqnarray}
-{\cal U} &=& C_1\; {\rm str}\left(
\xi_5^{(11)}\Sigma\;\xi_5^{(11)}\Sigma^\dagger
\right) \nonumber\\
&&+\frac{C_3}2 \sum_\nu {\rm str}\left(
\xi_\nu^{(11)}\Sigma\;\xi_\nu^{(11)}\Sigma + {\rm h.c.}
\right) \nonumber\\
&&+\frac{C_4}2 \sum_\nu {\rm str}\left(
\xi_{\nu5}^{(11)}\Sigma\;\xi_{5\nu}^{(11)}\Sigma + {\rm h.c.}
\right) \nonumber\\
&&+C_6\; \sum_{\mu<\nu} {\rm str}\left(
\xi_{\mu\nu}^{(11)}\Sigma\;\xi_{\nu\mu}^{(11)}\Sigma^\dagger
\right) ,
\label{eq:U}
\end{eqnarray}
where (following the notation of Ref.~\cite{VandeWater:2005uq})
$\xi_B^{(n)}$ is a $4n\times 4n$ matrix with the $4\times 4$
taste matrix $\xi_B$ repeated along the diagonal $n$ times.
This potential contributes, along with the mass term,
to pseudo-Goldstone particle masses, whose LO form (for flavor off-diagonal
states) is exemplified by
\begin{equation}
m_{xy,B}^2 = B_0 (m_x+m_y) + a^2 \Delta(\xi_B).
\end{equation}
Here $m_{xy,B}$ is the mass of the pseudo-Goldstone boson composed of a valence
quark of mass $m_x$ and a valence antiquark of mass $m_y$, and having taste $B$.
The taste-dependent discretization errors $\Delta(\xi_B)$ depend
on the LECs $C_1$, $C_3$, $C_4$ and $C_6$---explicit forms are given
in Ref.~\cite{Aubin:2003mg}.
In addition ${\cal U}$ leads to four-pion vertices which contribute to
the desired $B-$parameters at one-loop order.

Both the $m_0$ term and the two strace potential ${\cal U}'$
lead to ``hairpin'' (quark-line disconnected) vertices. 
Only the former contributes to the diagrams that enter here, and thus
we do not reproduce the form of ${\cal U}'$.

The procedure for accounting for rooting has been explained
in Ref.~\cite{Aubin:2003mg}. In essence, one must include by
hand a factor of $1/4$ for each contribution which corresponds to
a sea-quark loop. 
In the present calculation, it turns
out that there are no diagrams containing sea-quark loops, 
as explained in Sec.~\ref{subsec:NLOSU3SChPT}. The only place
where the $1/4$'s enter is in the quark loops implicitly contained
in quark-line disconnected meson propagators. The impact of the
$1/4$'s is worked out in Ref.~\cite{Aubin:2003mg}.
We quote the result only for the taste-singlet channel,
since this is the only disconnected propagator we need.
For a valence meson composed of a quark and its antiquark having
mass $m_x$ converting to a similar meson composed of quark and antiquark
of mass $m_y$, the disconnected part of the propagator is
(after sending $m_0\to\infty$)
\begin{eqnarray}
D_{xy}^I(q) = - \frac43 \frac{(q^2+U_I) (q^2 + D_I) (q^2+S_I)}
{(q^2+X_I)(q^2+Y_I) (q^2+\pi^0_I) (q^2+\eta_I)}
.
\label{eq:DI}
\end{eqnarray}
Here we are using the same compact notation used to present
results for $B_K$ in Ref.~\cite{Bae:2010ki}.
$X_I$ is the squared mass of the flavor off-diagonal, taste
singlet pion created by $\bar D_1 (\gamma_5\otimes \xi_I) D_2$,
which at LO is 
\begin{equation}
X_I = 2 B_0 m_x + a^2 \Delta(\xi_I).
\end{equation}
$Y_I$, $U_I$, $D_I$ and $S_I$ are defined similarly.
By contrast, $\pi^0_I$ and $\eta_I$ are the mass eigenstates in the
sea-quark sector, which thus include hairpin contributions, and
are given by
\begin{equation}
\frac{U_I\!+\!D_I\!+\!S_I}3 \mp \frac13 \sqrt{U_I^2\!+\!D_I^2\!+\!S_I^2\!-\!
U_ID_I\!-\!U_IS_I\!-\!D_IS_I}
,
\label{eq:pietaI}
\end{equation}
with the upper sign for the $\pi^0_I$ and the lower for the $\eta_I$.
In the isospin limit $m_u=m_d$ one recovers the familiar results
$\pi^0_I=U_I$ and $\eta_I=(U_I+2S_I)/3$.

The final issue to be discussed is the impact
of using a mixed action, with different types of staggered
valence and sea quarks. 
Here we can rely on the corresponding analysis for $B_K$~\cite{Bae:2010ki}.
The conclusion is that mixed-action effects can enter either
through loops involving mixed valence-sea pions or through the
presence of new hairpin vertices of vector and axial taste.
As will be seen in Sec.~\ref{subsec:NLOSU3SChPT}, however,
in the present calculation there are, at one-loop, no contributions
from mixed pions, and no contributions from vector and axial hairpins.
Thus the only impact of using a mixed action is that the values of
the LECs associated with discretization errors are changed. 
This is not a concern, however, since
these values are to be determined by fits to simulation results.

\subsection{Operator mapping at leading order}
\label{sec:mapLO}

In this subsection we map the BSM operators used
in lattice calculations, i.e. the
${\cal O}_j^{{\rm Lat,Actual}}$ of Eq.~(\ref{eq:Ojlaterror}),
into the chiral effective theory at leading order (LO).
We must first map these operators into the Symanzik effective
action. This is simplified by working at LO,
which implies that we can drop corrections 
proportional to $\alpha/(4\pi)$, $\alpha^2$ and $a^2$.
It then follows from Eq.~(\ref{eq:Ojlaterror}) that
the LO mapping is simply into the ${\cal O}_j^{{\rm Cont}'}$.
Thus our task is to map the latter operators in the 
Symanzik effective theory into the chiral Lagrangian.

The method for doing so was developed in
Ref.~\cite{Lee:1999zxa} and used for the $B_K$ operator in 
Ref.~\cite{VandeWater:2005uq}.
One introduces spurion fields in such a way that quark-level
operators become invariant under chiral transformations,
then determines the LO operators in the chiral effective theory
containing these spurions.
Since all that matters are chiral transformation properties,
the choice of color contraction is irrelevant, so both
${\cal O}_2^{{\rm Cont}'}$ and ${\cal O}_3^{{\rm Cont}'}$ 
map into the same set of LO operators (of course with different LECs).
The same statement holds for ${\cal O}_4^{{\rm Cont}'}$ and ${\cal O}_5^{{\rm Cont}'}$.

It turns out to be simplest to map the component parts of
the ${\cal O}_j^{{\rm Cont}'}$ separately
[see Eqs.~(\ref{eq:O2cont'}-\ref{eq:O5cont'})]. We begin with
\begin{equation}
\mathcal{O}_{S+P} = 2\left[ \mathcal{O}^{{\rm Cont}'}_{S} +
 \mathcal{O}^{{\rm Cont}'}_{P} \right]
,
\end{equation}
where the factor of $2$ is for later convenience.
We do not specify the color contraction since the
subsequent results hold for both choices.
This operator is a component of 
${\cal O}_2^{{\rm Cont}'}$ and ${\cal O}_3^{{\rm Cont}'}$.
We first rewrite it in generic form
\begin{eqnarray}
\mathcal{O}_{S+P} &=& \bar Q_R (1\otimes F_{1L}) Q_L\;
\bar Q_R (1\otimes F_{2L}) Q_L 
\nonumber\\
&&+
\bar Q_L (1\otimes F_{1R}) Q_R\;
\bar Q_L (1\otimes F_{2R}) Q_R 
\label{eq:QformS+P}
\end{eqnarray}
where the second term on the right-hand side is the parity conjugate (p.c.)
of the first.
In the matrices $(1\otimes F)$ the ``$1$'' indicates the
identity matrix in spin space, while the spurion $F$
is a matrix in the tensor product of flavor and taste space.
In order to obtain the desired operator, one must choose the
spurions as follows:
\begin{equation}
F_{1L},F_{1R} \to \delta_{i,S1}\delta_{D1,j}\; \xi_5 
\ {\rm and} \
F_{2L},F_{2R} \to \delta_{i,S2}\delta_{D2,j}\; \xi_5 ,
\label{eq:setspurions}
\end{equation}
where $i$ and $j$ are flavor indices.

Next we note that $\mathcal{O}_{S+P}$ is invariant under
chiral transformations if the spurions transform as
\begin{equation}
F_{kL}\to R F_{kL} L^\dagger 
\quad {\rm and}\quad
F_{kR}\to L F_{kR} R^\dagger 
,
\label{eq:Ftransform}
\end{equation}
where $k=1,2$.
We now imagine that $\mathcal{O}_{S+P}$ is inserted into
the Symanzik action.
Then the desired matrix elements 
can be obtained by taking (a sum of) functional derivatives
with respect to appropriate elements of the $F_{kL}$ and $F_{kR}$
and then setting the spurions to zero.

The final step is to add to the chiral Lagrangian
all operators composed of the new spurions together
with $\Sigma$, $\Sigma^\dagger$, $M$, $M^\dagger$, derivatives
and  spurions coming from the ${\cal O}(a^2)$ terms in the Symanzik action,
such that the overall
operator is invariant under chiral transformations, Euclidean rotations,
and parity.
Furthermore, we need only keep operators proportional to $F_{1L}F_{2L}$
and $F_{1R}F_{2R}$, since only these will survive when we
take the functional derivatives of the chiral partition function
needed to obtain the desired matrix elements.
We see immediately that the LO operators will be those
involving $\Sigma$ and $\Sigma^\dagger$ alone,
with no derivatives, mass terms or $a^2$ spurions.
There are two such operators:
\begin{equation}
{\cal O}^\chi_a
= {\rm str}\left(\Sigma F_{1L} \Sigma F_{2L}\right) + {\rm p.c.}
\label{eq:Ochia}
\end{equation}
where the parity conjugate is obtained by $L\leftrightarrow R$
and $\Sigma\leftrightarrow \Sigma^\dagger$, and
\begin{equation}
{\cal O}^\chi_b =
{\rm str}\left(\Sigma F_{1L}\right){\rm str}\left(\Sigma F_{2L}\right) + 
{\rm p.c.}
\label{eq:Ochib}
\end{equation}
One now sets the spurions to their original values,
as in Eq.~(\ref{eq:setspurions}).
The resulting two operators will appear with independent,
unknown coefficients.

To map ${\cal O}_2^{{\rm Cont}'}$ and ${\cal O}_3^{{\rm Cont}'}$
into the chiral theory we also need to consider
$\mathcal{O}^{{\rm Cont}'}_{T}$.
It turns out that this operator maps into the same
two chiral operators as $\mathcal{O}_{S+P}$.
To see this, we note that, in addition to the form
(\ref{eq:OTII}), the operator can be written 
\begin{equation}
\mathcal{O}^{{\rm Cont}'}_{T}
= \sum_{\mu<\nu} 
\bar S_1 (\gamma_\mu\gamma_\nu\gamma_5 \otimes \xi_5) D_1\;
\bar S_2 (\gamma_\mu\gamma_\nu\gamma_5 \otimes \xi_5) D_2 
.
\end{equation}
Combining these two forms one finds that the spurion representation
of the operator is
\begin{equation}
\mathcal{O}^{{\rm Cont}'}_{T} = 
\bar Q_R (\gamma_\mu\gamma_\nu\otimes F_{1L}) Q_L\;
\bar Q_R (\gamma_\mu\gamma_\nu\otimes F_{2L}) Q_L 
+ {\rm p.c.}
\label{eq:QformT}
\end{equation}
From the point of view of chiral symmetry, this operator
transforms in exactly the same way as
${\cal O}_{S+P}$, Eq.~(\ref{eq:QformS+P}).
Thus its mapping into chiral operators has the same form.

The final stage of the mapping is to note that the
relative coefficient of the two chiral operators 
${\cal O}^\chi_a$ and ${\cal O}^\chi_b$ is fixed,
so that there is only one overall LEC.
This holds only for the particular
linear combinations of ${\cal O}_{S+P}$ and
${\cal O}_T^{{\rm Cont}'}$ that appear in
${\cal O}_{2,3}^{{\rm Cont}'}$.
The key observation is that the coefficients of the
two chiral operators are exactly the same as they would be
if one set $a=0$ in the Symanzik theory. This is because
all factors of $a$ are explicit, and there are none in either
chiral operator. But setting $a=0$ in the Symanzik theory leads
to the PQ continuum theory considered in Sec.~\ref{sec:Olat_cont}.
Since taste symmetry is exact in this theory,
the results (\ref{eq:tasteI}) and (\ref{eq:FierzO}) 
must hold for the matrix elements at
both the quark level and the chiral level.
Furthermore, they must hold order by order in the momentum--quark-mass expansion of SChPT, and
in particular they must hold at LO in the standard power counting.
The leading-order matrix elements are simple to calculate.
If we write the chiral mapping of ${\cal O}_{2}^{{\rm Cont}'}$ as
\begin{equation}
{\cal O}^\chi_2 = 
\left(c_{2a} {\cal O}^\chi_a + c_{2b} {\cal O}^\chi_b\right)
,
\end{equation}
then we find
\begin{eqnarray}
\langle\overline{K}_{B1}^{0}|{\cal O}^\chi_2|K_{B2}^{0}\rangle_{\rm LO}
&=& -\frac{32}{f^2} c_{2b} \delta_{B,P}
\label{eq:Ochi2LOb}
\\
\langle\overline{K}_{B21}^{0}|{\cal O}^\chi_2|K_{B12}^{0}\rangle_{\rm LO}
&=&
-\frac{8}{f^2} c_{2a} \frac14 {\rm tr}(\xi_B\xi_5\xi_B\xi_5)
.
\label{eq:Ochi2LOa}
\end{eqnarray}
These results are consistent with (\ref{eq:tasteI}) and (\ref{eq:FierzO}) 
only if $c_{2a}=c_{2b}$.
We thus conclude that
\begin{equation}
{\cal O}^\chi_2 = 
c_2 \left({\cal O}^\chi_a + {\cal O}^\chi_b\right)
.
\label{eq:Ochi23}
\end{equation}
The same form holds for ${\cal O}^\chi_3$ but with a
different coefficient, $c_3$.

The fact that, at LO, there is only one unknown LEC
could have been anticipated from the result that there
is only a single LEC in the mapping of $O_2$ into
continuum ChPT~\cite{Becirevic:2004qd}.
We also note that a similar analysis holds for
the chiral mapping of the $B_K$ operator~\cite{VandeWater:2005uq}.

We now turn to the chiral mapping of ${\cal O}_{4,5}^{{\rm Cont}'}$.
These are composed of
\begin{equation}
\mathcal{O}_{S-P} = 2\left[ \mathcal{O}^{{\rm Cont}'}_{S} -
 \mathcal{O}^{{\rm Cont}'}_{P} \right]
,
\end{equation}
and
\begin{equation}
\mathcal{O}_{V-A} = 2\left[ \mathcal{O}^{{\rm Cont}'}_{V} -
 \mathcal{O}^{{\rm Cont}'}_{A} \right]
.
\end{equation}
In terms of spurions, the former operator is
\begin{equation}
\mathcal{O}_{S-P} = \bar Q_R (1\otimes F_{1L}) Q_L\;
\bar Q_L (1\otimes F_{2R}) Q_R + {\rm p.c.}
\label{eq:QformS-P}
\end{equation}
Here, the spurions $F_{1L}$, $F_{2R}$ and their parity conjugates
transform as above [Eq.~(\ref{eq:Ftransform})], and are set at the end to
the same values as in (\ref{eq:setspurions}). Note, however,
that Eq.~(\ref{eq:QformS-P}) differs from the spurion form
of ${\cal O}_{S+P}$, Eq.~(\ref{eq:QformS+P}).
The former is proportional to $F_L F_R$, while the latter 
to $F_L F_L$.
This leads to the presence of only a single LO chiral operator,
\begin{equation}
{\cal O}^\chi_c =
{\rm str}\left(\Sigma F_{1L}\right){\rm str}
\left(\Sigma^\dagger F_{2R}\right) + 
{\rm p.c.}
\end{equation}

Turning now to the $V-A$ operator,
its spurion form is
\begin{equation}
\mathcal{O}_{V-A} = \sum_\mu 
\bar Q_L (\gamma_\mu\otimes \tilde F_{1L}) Q_L\;
\bar Q_R (\gamma_\mu\otimes \tilde F_{2R}) Q_R + {\rm p.c.},
\label{eq:OV-A}
\end{equation}
where the spurions now transform as
\begin{equation}
\tilde F_{kL}\to L F_{kL} L^\dagger 
\quad {\rm and}\quad
\tilde F_{kR}\to R F_{kR} R^\dagger 
.
\label{eq:tildeFtransform}
\end{equation}
At the end they are set to the same values as the other
spurions,
\begin{equation}
\tilde F_{1L},\tilde F_{1R} \to \delta_{i,S1}\delta_{D1,j}\; \xi_5 
\ {\rm and} \ 
\tilde F_{2L},\tilde F_{2R} \to \delta_{i,S2}\delta_{D2,j}\; \xi_5 
.
\end{equation}
The single LO chiral operator that this maps to is
\begin{equation}
{\cal O}^\chi_d
= {\rm str}\left(\Sigma^\dagger F_{1L} 
                 \Sigma F_{2R}\right) + {\rm p.c.}
\end{equation}

Combining these two operators into ${\cal O}_{4,5}^{{\rm Cont}'}$
and enforcing the relations
(\ref{eq:tasteI}) and (\ref{eq:FierzO}) from taste symmetry and Fierzing, we find again that
the coefficients are related.
The chiral mapping is to
\begin{equation}
{\cal O}^\chi_{4,5} = 
c_{4,5} \left({\cal O}^\chi_c + {\cal O}^\chi_d\right)
.
\label{eq:Ochi45}
\end{equation}

Finally, we need to map the pseudoscalar densities appearing in
the denominator of Eq.~(\ref{eq:Bjlatt}) into the chiral theory.
Since we are working at LO, the operators in the Symanzik effective theory
that we need to map are
$[\bar S_k(\gamma_5\otimes \xi_5) D_k]^{{\rm Cont}'}$ for $k=1,2$.
This is a standard exercise and we find
\begin{eqnarray}
\left[\bar S_k(\gamma_5\otimes \xi_5) D_k\right]^{{\rm Cont}'}
&\longrightarrow&
c_{\rm bil} {\cal O}^\chi_{{\rm bil},k}
\\
{\cal O}^\chi_{{\rm bil},k} &=& 
{\rm str}\left(\Sigma F_{kL}\right) - {\rm p.c.}
\label{eq:Ochibil}
\end{eqnarray}
Expressing the constant $c_{\rm bil}$ in terms of other LECs
is not useful here since the 
corresponding constants in the mapping
of the numerator of (\ref{eq:Bjlatt}) are unknown.

\subsection{Mapping at next-to-leading order}
\label{subsec:NLO}

An important conclusion from the previous subsection is
that, because the LO chiral operators ${\cal O}^\chi_j$
contain no derivatives, they give rise to LO matrix elements
that are non-vanishing in the chiral limit.
This is seen explicitly in Eqs.~(\ref{eq:Ochi2LOb}) and (\ref{eq:Ochi2LOa}).
Unlike for the $B_K$ operator, there is no chiral suppression, a result
that is well known in continuum phenomenology.
This means that higher-order chiral operators which contain factors
of $M$, $a^2$, $\alpha/(4\pi)$, $\alpha^2$, or which contain derivatives
can only give rise to analytic contributions to the $B_j$.
Non-analytic contributions at NLO can arise only from one-loop
diagrams involving the LO chiral operators.

Because of this, 
we can determine the functional form of the NLO contributions
to the $B_j$ without explicitly enumerating
all the higher-order chiral operators which appear when we
map the ${\cal O}_j^{\rm Lat, Actual}$ into the chiral theory. 
We know that operators which come with two
derivatives will lead to analytic terms $\propto m_K^2$,
while operators arising from discretization errors in the action
or the operators lead to analytic terms $\propto a^2$. 
This holds for both the numerators and
denominators of the $B_j$.

We can also see that there are no NLO analytic terms $\propto \alpha/(4\pi)$.
These only arise 
from the ``other taste'' operators of Eq.~(\ref{eq:Ojlaterror}),
which enter in the numerators of the $B_j$.
When we match these four-fermion operators from the lattice
to the Symanzik effective theory, the LO continuum
operators that result will still have tastes other than $P$
(since taste-breaking would bring in further factors of $a_\alpha^2$).
Thus their matrix elements with taste $P$ external kaons 
vanish at LO. This in turn implies that the LO chiral representation
of these operators (which contain no derivatives) must have vanishing
tree-level matrix elements between kaons of taste $P$.
Their one-loop matrix elements will be non-vanishing,
but, because of the overall factor of $\alpha/(4\pi)$,
these contributions enter at next-to-next-to-leading order. 

Finally, we consider terms involving insertions of the quark
mass matrix $M$. Here an explicit enumeration is useful.
We find two types of chiral operators.
First, those in which the LO operators are multiplied by
${\rm str}(M\Sigma^\dagger) + {\rm p.c.}$  These lead to
analytic corrections $\propto m_u+m_d+m_s$.
Second, factors of $M\Sigma^\dagger$ and $\Sigma M^\dagger$
can be inserted in the LO chiral operators.
The corrections to ${\cal O}^\chi_{2,3}$ are, for example,
\begin{eqnarray}
&&{\rm str}\left(\Sigma F_{1L}\Sigma F_{2L}\Sigma M^\dagger\right) 
+ (1\leftrightarrow 2)
+ {\rm p.c.}
\\
&&{\rm str}\left(\Sigma F_{1L}\Sigma M^\dagger\right) 
{\rm str}\left(\Sigma F_{2L}\right) + (1\leftrightarrow 2)
+ {\rm p.c.}
\end{eqnarray}
The first of these operators does not contribute to the
desired matrix element at tree-level,
while the second gives a contribution $\propto m_x+m_y$.
There are no terms $\propto m_x-m_y$,
as can also be seen more generally using CPS symmetry~\cite{Bernard:1985wf}.
A similar analysis leads to the same conclusion for
the corrections to ${\cal O}^\chi_{4,5}$.

In summary, NLO analytic corrections to both the numerators and denominators
of the $B_j$ are proportional to $m_K^2$, $a^2$, $a_\alpha^2$, $\alpha^2$,
$m_u+m_d+m_s$ and $m_x+m_y$. Since we use pseudo-Goldstone external kaons,
for which $m_K^2\propto m_x+m_y$ at LO (with no $a^2$ terms), the
$m_K^2$ and $m_x+m_y$ corrections can be combined into a single term.

\section{SChPT results for $B-$parameters at NLO}
\label{sec:results}

The analysis of the previous section shows that, to
NLO in SChPT, we have
\begin{eqnarray}
B_j &=& \frac{2}{N_j}
\frac{\langle\overline{K}_{P1}^{0}|{\cal O}^\chi_j|K_{P2}^{0}\rangle_{\rm 1-loop}}
{\langle\overline{K}_{P1}^0|{\cal O}^\chi_{{\rm bil},1}|0\rangle_{\rm 1-loop}
\langle0|{\cal O}^\chi_{{\rm bil},2}|{K}_{P2}^0\rangle_{\rm 1-loop}}
\nonumber\\
&& + {\rm analytic\ NLO} ,
\label{eq:BjNLO}
\end{eqnarray}
with the chiral operators in the numerator 
defined in Eqs.~(\ref{eq:Ochi23}) and (\ref{eq:Ochi45}),
while those in the denominator are given in Eq.~(\ref{eq:Ochibil}).
The subscript ``1-loop'' here means the {\em sum} of 
tree-level and one-loop contributions.
At LO, the matrix elements in both numerator and denominator are constants,
independent of quark masses and kaon momenta. Explicitly, we find
\begin{equation}
B_j^{\rm LO} 
= \frac{2}{N_j} \frac{\mp c_j 32/f^2}{(8ic_{\rm bil}/f)^2}
= \pm \frac{c_j}{N_j c_{\rm bil}^2} 
\,.
\label{eq:BjLO}
\end{equation}
Throughout this section the upper sign applies for
$j=2,3$, and the lower sign, for $j=4,5$.
The result (\ref{eq:BjLO})
has no predictive power since we do not know the constants
$c_j$. The $B_j^{\rm LO}$ are simply the
values of the $B_j$ in the joint chiral-continuum limit.

The predictive power of Eq.~(\ref{eq:BjNLO}) arises because
the one-loop contributions involve only the LO chiral operators,
implying that the {\em relative}
contribution of the chiral logarithms is determined.

\begin{figure}[htbp!]
\subfigure[]{\includegraphics[angle=270,width=0.23\textwidth]{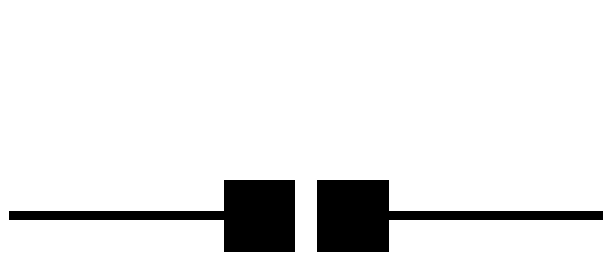}}
\subfigure[]{\includegraphics[angle=270,width=0.23\textwidth]{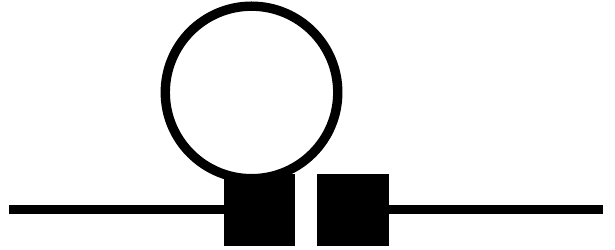}}
\subfigure[]{\includegraphics[angle=270,width=0.23\textwidth]{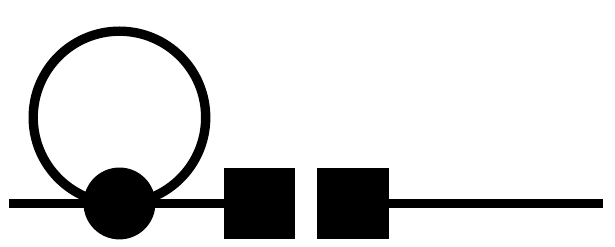}}
\subfigure[]{\includegraphics[angle=270,width=0.23\textwidth]{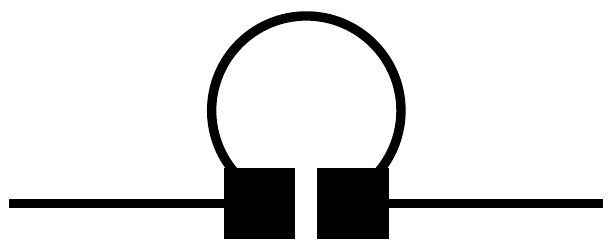}}
\subfigure[]{\includegraphics[angle=270,width=0.23\textwidth]{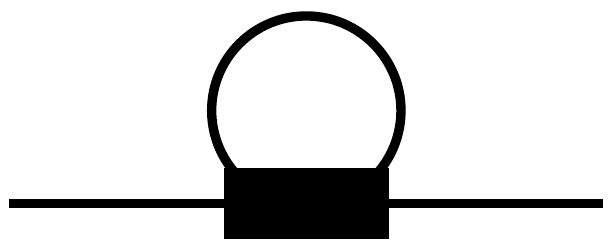}}
\subfigure[]{\includegraphics[angle=270,width=0.23\textwidth]{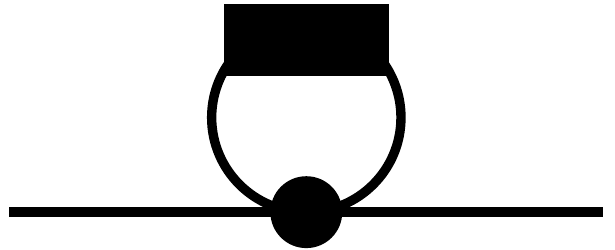}}
\caption{SChPT diagrams contributing to the numerator of $B_K$:
(a) tree-level; (b-f) one-loop. The double strace components
of the chiral operators are represented by two square boxes,
one per strace, while the single strace components are shown
with one rectangular box. 
(It turns out that there is no contribution from a diagram of the
form of (f) but with a two strace operator.)
The filled circle is the full LO
vertex from the SChPT Lagrangian, including ${\cal O}(a^2)$ terms.
For (b) and (c) we have not shown separately the diagrams in which
the loop is attached to the other external kaon. 
\label{fig:BKnum}}
\end{figure}

The tree-level and one-loop contributions to the kaon matrix
elements in the numerator of $B_j$ are shown in Fig.~\ref{fig:BKnum}.
Here we distinguish between the single and double strace sub-operators
contained in the ${\cal O}^\chi_{j}$. For example,
from Eq.~(\ref{eq:Ochi23}) we see that the single strace component of
${\cal O}^\chi_2$ is ${\cal O}^\chi_a$ of Eq.~(\ref{eq:Ochia})
while the double strace component is
${\cal O}^\chi_b$ of Eq.~(\ref{eq:Ochib}).
Given the tastes of the external kaons, 
only the double strace components contribute at tree-level,
as shown for the case of ${\cal O}^\chi_2$ by Eq.~(\ref{eq:Ochi2LOa}).
Both components contribute at one-loop order, as shown in
Fig.~\ref{fig:BKnum}.

\begin{figure}[htbp!]
\subfigure[]{\includegraphics[angle=270,width=0.13\textwidth]{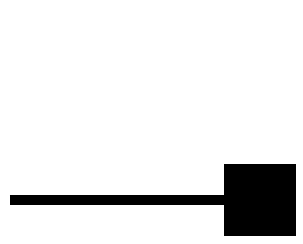}}
\subfigure[]{\includegraphics[angle=270,width=0.14\textwidth]{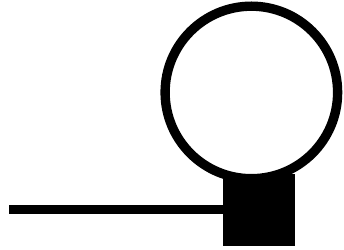}}
\subfigure[]{\includegraphics[angle=270,width=0.14\textwidth]{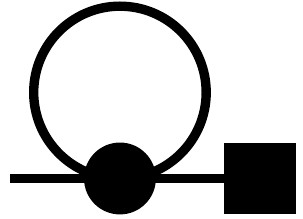}}
\caption{SChPT diagrams contributing to the single
kaon matrix elements in the denominator of $B_K$:
(a) tree-level; (b-c) one-loop. Notation is as in
Fig.~\protect\ref{fig:BKnum}.
\label{fig:BKden}}
\end{figure}

The contributions to the denominator are shown in Fig.~\ref{fig:BKden}.
They are simpler since there is only a single strace component.
Noting that the square boxes in Figs.~\ref{fig:BKnum} and \ref{fig:BKden}
correspond to identical chiral operators, and accounting for the fact that
the loops in Figs.~\ref{fig:BKnum}(b) and (c) can appear on either external
kaon line, we see that the contributions to $B_j$ from
Figs.~\ref{fig:BKnum}(b) and (c) cancel with those from
Figs.~\ref{fig:BKden}(b) and (c). Wavefunction renormalization factors
also cancel. These are the same cancellations as those found for
$B_K$ in Ref.~\cite{VandeWater:2005uq}. Thus we need only keep the
diagrams of Figs.~\ref{fig:BKnum}(d), (e) and (f).

\begin{figure}[htbp!]
\subfigure[]{\includegraphics[angle=270,width=0.23\textwidth]{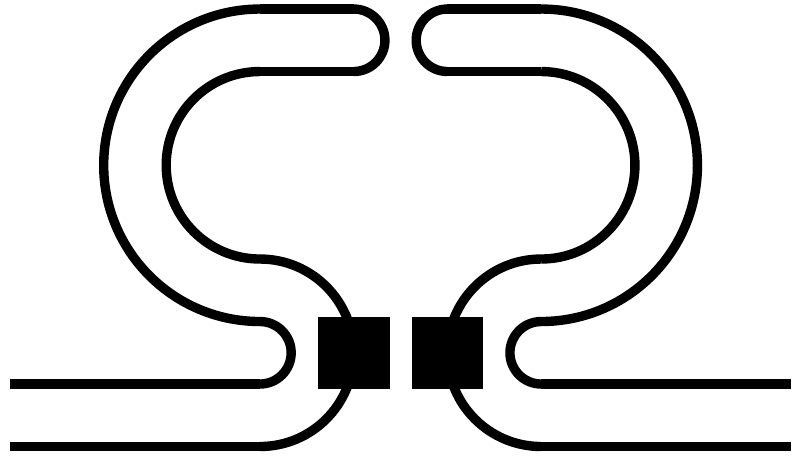}}
\subfigure[]{\includegraphics[angle=270,width=0.46\textwidth]{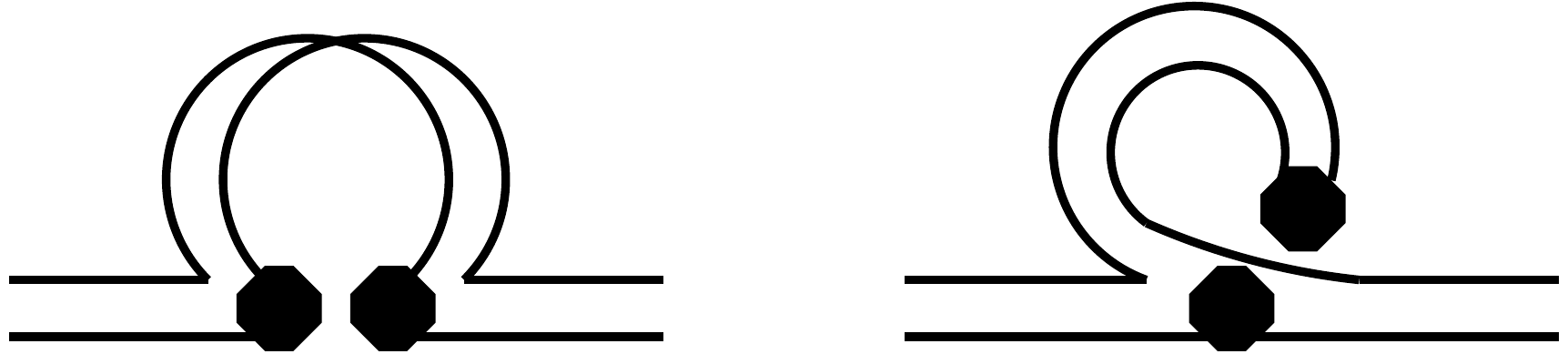}}
\subfigure[]{\includegraphics[angle=270,width=0.23\textwidth]{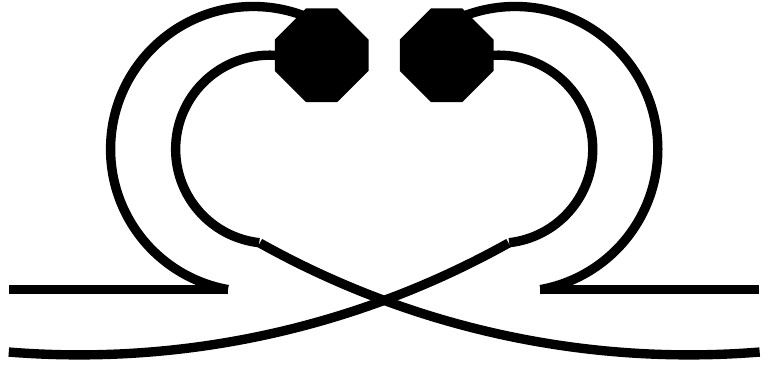}}
\caption{Quark-line diagrams contributing to
(a) Fig.~\protect\ref{fig:BKnum}(d), 
(b) Fig.~\protect\ref{fig:BKnum}(e), and 
(c) Fig.~\protect\ref{fig:BKnum}(f). 
In each case, each of the filled boxes correspond to one of
the factors of $\Sigma$ (or $\Sigma^\dagger$) in the chiral operators.
The square boxes arise from the double strace operators and thus
have flavor $\bar S_1D_1$ or $\bar S_2 D_2$.
The octagonal boxes arise from the single strace operators and
thus have flavor $\bar S_1 D_2$ or $\bar S_2 D_1$.
\label{fig:quarkline}}
\end{figure}

It is useful to draw the quark-line diagrams contributing to
Figs.~\ref{fig:BKnum}(d-f). These are shown in Fig.~\ref{fig:quarkline}.
We recall that these are primarily a device for tracking the flavor indices 
of mesons in the diagrams that contribute to the SChPT calculation.
They also correspond, however, to different ways of routing the
quark propagators of the underlying lattice calculation
so as to make loop diagrams. In the latter interpretation, each of
the boxes corresponds to one of the component bilinears in the
four-fermion operator, and in the case where the boxes are octagons
rather than squares one must first Fierz transform the operator
into its $(\bar S_1 D_2) (\bar S_2 D_1)$ form.

We see from Fig.~\ref{fig:quarkline} that there are no
diagrams involving valence-sea mesons. 
Such mesons do contribute to Figs.~\ref{fig:BKnum}(b) and (c)
and to Figs.~\ref{fig:BKden}(b) and (c), but these
contributions cancel at NLO as discussed above.
This is an important simplification because it means that we
do not need to determine the masses of valence-sea mesons,
which contain a different $a^2$ contribution from the valence-valence
mesons. Thus one of the potential complications from using a mixed
action does not occur.

We also see, from Fig.~\ref{fig:quarkline}(a),
that Fig.~\ref{fig:BKnum}(d)
involves only the quark-disconnected, hairpin part of the
meson propagator. This holds because one cannot have
a quark-connected propagator joining mesons composed of quarks having
different flavors ($\bar S_2 D_2$ versus $\bar S_1 D_1$).
Furthermore, because both the external kaon and the bilinear represented by
the square box have taste $P$, the meson in the loop must be a taste
singlet. An important corollary is that the second
complication due to the use of a mixed-action---namely, 
the change in taste-$V$ and taste-$A$ hairpin vertices---does not 
impact the present calculation.

\subsection{NLO SU(3) SChPT result}
\label{subsec:NLOSU3SChPT}

In this section we give the general form of the
next-to-leading order corrections for a $1+1+1$ flavor theory,
i.e. the rooted theory in which we keep the sea quark
masses general.

We break up the corrections as follows 
\begin{equation}
B_j = B_{j}^{\rm LO}\left[1 + \delta B^{\rm anal}_j
+ \delta B^{\rm conn}_j + \delta B^{\rm disc}_j \right]
,
\label{eq:Bjfinal}
\end{equation}
where the first correction contains the analytic term,
the second is the contribution from Figs.~\ref{fig:BKnum}(e) and
(f), which we refer to as ``connected'' since they do not
involve hairpin vertices,
and the third is the contribution from Fig.~\ref{fig:BKnum}(d),
which is ``disconnected'' as it involves only hairpin vertices.

The analytic terms have been discussed in Sec.~\ref{subsec:NLO}, 
and have the form
\begin{eqnarray}
\delta B^{\rm anal}_j &=& c_{j1} (m_x+m_y)
+ c_{j2} (m_u+m_d+m_s) 
\nonumber\\
&&
+ c_{j3} a^2
+ c_{j4} a_\alpha^2
+ c_{j5} \alpha^2
.
\end{eqnarray}

We find the connected contributions to be
\begin{eqnarray}
\delta B^{\rm conn}_{j}
&=&
\frac{-1}{(4\pi f)^2}
\frac{1}{16} \sum_B
\Big[2\ell(K_B) - 2 K_P \tilde\ell(K_B)
\nonumber\\
&&
\pm \ell(X_B) \pm \ell(Y_B) \Big],
\end{eqnarray}
where $B$ is a taste label, which is summed over the 16 possibilities.
We use the following abbreviations for
meson mass squareds~\cite{Bae:2010ki}:
$X_B=m_{xx,B}^2$, $Y_B=m_{yy,B}^2$ and $K_B = m_{xy,B}^2$. 
The chiral logarithmic functions are, in infinite volume,
\begin{equation}
\ell(X) = X \ln \frac{X}{\mu^2},\ \
\tilde\ell(X) = - \ln \frac{X}{\mu^2} -1 ,
\end{equation}
where $\mu$ is the renormalization scale in dimensional
regularization. The dependence on $\mu$ is absorbed by
the implicit $\mu$ dependence of $c_{j1}$ and 
(for $j=4,5$) $c_{j4}$.
Finite volume corrections to these logarithms
are standard and are given, e.g., in Ref.~\cite{VandeWater:2005uq}.

We find the disconnected contributions to be
\begin{equation}
\delta B^{\rm disc}_j = \pm
\frac{-1}{4 f^2}
\int \frac{d^4q}{(2\pi)^4} 
\left[D_{xx}^I(q) + D_{yy}^I(q) + 2 D_{xy}^I(q)\right]
,
\label{eq:Bdiscsimple}
\end{equation}
where the disconnected propagators are given in Eq.~(\ref{eq:DI}).
It is straightforward, though tedious, to evaluate these
integrals by summing the contributions from the
poles in the propagators, following the method of 
Ref.~\cite{Sharpe:2001fh}. For the sake of brevity, however,
we only quote the result in the isospin symmetric limit
(i.e. for a $2+1$ flavor theory):
\begin{eqnarray}
\delta B_j^{\rm disc}&=& \pm
\frac{1}{3(4\pi f)^2} (I_\eta + I_X+I_Y )
\label{eq:Bdisc}
\end{eqnarray}
where
\begin{eqnarray}
I_\eta &=&
\ell(\eta_I) (L_I\!-\!\eta_I)(S_I\!-\!\eta_I)
\left[\frac1{X_I\!-\!\eta_I}+\frac1{Y_I\!-\!\eta_I}\right]^2
\label{eq:Ieta}
\end{eqnarray}
and
\begin{eqnarray}
\lefteqn{I_X =
\frac{(L_I\!-\!X_I)(S_I\!-\!X_I)}{(\eta_I\!-\!X_I)}
\Bigg\{\tilde\ell(X_I) + \ell(X_I)\times}
\label{eq:IX}
\\
&&
\left[\frac2{Y_I\!-\!X_I}+\frac1{L_I\!-\!X_I}
+\frac1{S_I\!-\!X_I}
-\frac1{\eta_I\!-\!X_I}\right]\Bigg\}
\nonumber\\
I_Y &=& I_X (X\leftrightarrow Y)
.
\label{eq:IY}
\end{eqnarray}
Despite appearances, this expression is finite
when $X_I \to \eta_I$, $L_I$, $S_I$ or $Y_I$,
and similarly for $Y_I$.
The $\ln\mu$ dependence is proportional to
\begin{align}
&L_I+S_I-\eta_I-X_I-Y_I \nonumber\\
&=\frac13(2 L_I+S_I) - (X_I+Y_I)
\\
&= B_0\left[\frac13(m_u\!+\!m_d\!+\!m_s) 
- (m_x\!+\!m_y)\right] 
- a^2 \Delta(\xi_I)
,
\end{align}
and can thus be absorbed by
shifts in $c_{j1}$, $c_{j2}$ and $c_{j4}$.

We can check our result with that of
Ref.~\cite{Becirevic:2004qd}
by taking the continuum, unquenched limit.
In this limit, all tastes are degenerate, and
\begin{eqnarray}
D_{xx}^I(q) &\longrightarrow& -2 \frac1{q^2+m_\pi^2}
+\frac23 \frac1{q^2+m_\eta^2},
\\
D_{xy}^I(q) &\longrightarrow& -\frac43 \frac1{q^2+m_\eta^2},
\\
D_{yy}^I(q) &\longrightarrow& -4 \frac1{q^2+ S}
+\frac83 \frac1{q^2+m_\eta^2}.
\end{eqnarray}
Thus
\begin{equation}
\delta B_j^{\rm conn}\to 
\frac{-1}{(4\pi f)^2} \left[
2\ell(K) - 2K\tilde\ell(K) \pm \ell(m_\pi^2) \pm \ell(S)
\right],
\end{equation}
while
\begin{equation}
\delta B_j^{\rm disc}\to \pm
\frac{-1}{(4\pi f)^2} \left[
-\frac{1}2 \ell(m_\pi^2)
- \ell(S)
+\frac16 \ell(m_\eta^2) 
\right].
\end{equation}
Combining these results one finds that the
unphysical logarithms $\ell(S)$ cancel, and
the results agree with those of Ref.~\cite{Becirevic:2004qd}.
 
\subsection{SU(2) SChPT}

The utility of SU(3) ChPT at the
physical kaon mass is unclear, given the
relatively large value of the expansion parameter
$m_K^2/\Lambda_\chi^2$ ($\Lambda_\chi\sim 1\;$GeV).
Thus it has proved very useful to consider the
strange quark as heavy and work instead with SU(2) ChPT.
In this limit only pion loops lead to chiral logarithms,
while the kaon is a static source.
SU(2) ChPT was first developed in the continuum 
in Ref.~\cite{Roessl:1999iu},
and extended and applied to $B_K$ (and other quantities)
in Ref.~\cite{rbc-uk-08}.
The extension to staggered ChPT was described in Ref.~\cite{Bae:2010ki}
in the context of the application to $B_K$.
Our present application is very similar to (and indeed somewhat
simpler than) $B_K$, and thus we can take over much of
the work of Ref.~\cite{Bae:2010ki}.

We first make clear the limit that we are considering
in our extended partially quenched set-up.
We take the valence and sea strange-quark masses
($m_y$ and $m_s$, respectively) to be heavy, while $m_x$, $m_u$ and
$m_d$ remain light. All particles containing one or two strange
quarks of either type are treated as heavy, i.e. the valence
and sea quark kaons, $\bar y y$ particles and the $\eta_I$.
The masses of these heavy particles are considered to be
of the same order as $\Lambda_\chi$.

It is argued in Ref.~\cite{Bae:2010ki} that for $B_K$
one can obtain the general NLO SU(2) SChPT result from the 
NLO SU(3) SChPT result using the following recipe:
take the limit $m_\pi\ll m_K$
and treat $(m_\pi/m_K)^2$ as a small parameter
of size $(p/\Lambda_\chi)^2$,
drop chiral logarithms of heavy particles,
and replace all LECs with (unknown) functions of $m_y$ and $m_s$.
The only loops that remain are those of mesons containing
light quarks alone, e.g. $\bar x x$ particles and sea-quark pions.\footnote{%
$\eta_B$ particles for $B\ne I$ are light in the SU(2) limit,
but these do not appear in the expressions for the $B_j$.
The same is true for mixed mesons composed of a light sea quark
and a light valence antiquark, or vice-versa.}
The argument holds also for the present application.
Indeed, the argumentation is simpler here because the LO matrix
elements are not chirally suppressed. In light of this
simplicity, and because the treatment of Ref.~\cite{Bae:2010ki}
suppressed some details, we explain the argument here.

First, however, we give the result, as this will
facilitate the subsequent explanation.
Here, the full $1+1+1$ flavor result is simple enough 
to present.
It has the same form as the SU(3) result, Eq.~(\ref{eq:Bjfinal}),
\begin{eqnarray}
B_j &=& B_{j}^{\rm LO,SU(2)}\Bigg[1 + \delta B^{\rm anal,SU(2)}_j
\nonumber\\
&&
+ \delta B^{\rm conn,SU(2)}_j + \delta B^{\rm disc,SU(2)}_j \Bigg]
,
\label{eq:BjfinalSU2}
\end{eqnarray}
but now the overall constant, $B_j^{\rm LO,SU(2)}$,
is the value of $B_j$ in the SU(2) chiral limit
(and for $a=0$) rather than the SU(3) limit which
applies for $B_j^{\rm LO}$. $B_j^{\rm LO,SU(2)}$
thus has an unknown and general dependence on $m_y$ and $m_s$.
The analytic term becomes
\begin{eqnarray}
\delta B^{\rm anal,SU(2)}_j &=&
  d_{j1} m_x
+ d_{j2} (m_u+m_d)
\nonumber\\
&&+ d_{j3} a^2
+ d_{j4} a_\alpha^2
+ d_{j5} \alpha^2
,
\label{eq:deltaBanalSU(2)}
\end{eqnarray}
with new LECs.
Because only the light meson loop survives,
the connected contribution simplifies to
\begin{equation}
\delta B^{\rm conn,SU(2)}_{j}
=
\pm \frac{-1}{(4\pi f_2)^2}
\frac{1}{16} \sum_B \ell(X_B) .
\label{eq:BconnSU2}
\end{equation}
Here $f_2$ is the decay constant in the SU(2)
chiral limit.
Finally, the disconnected contribution 
can be obtained from Eq.~(\ref{eq:Bdiscsimple})
by noting that the light-particle poles
in $D_{xy}^I$ come with residues having an additional
suppression factor $m_{u,d}/m_{s,y}$, while those in
$D_{yy}^I$ (which are only present in a $1+1+1$ flavor
theory) are even more suppressed. Thus only
the $D_{xx}^I$ term contributes, and using
$S^I/\eta^I=3/2$ and $2\pi_I=U_I+D_I$
[which follow from Eq.~(\ref{eq:pietaI}) for $m_{s}\gg m_{u,d}$],
we find
\begin{eqnarray}
\delta B^{\rm disc,SU(2)}_{j}
&=&
\pm \frac{1}{2(4\pi f_2)^2}
\Bigg\{
\tilde\ell(X_I)\frac{(U_I\!-\!X_I)(D_I\!-\!X_I)}{(\pi_I\!-\!X_I)}
\nonumber\\
&&+
\ell(X_I)\left[2 -
\frac{(U_I\!-\!X_I)(D_I\!-\!X_I)}{(\pi_I\!-\!X_I)^2}\right]
\nonumber\\
&&+
\ell(\pi_I)
\frac{(U_I\!-\!\pi_I)(D_I\!-\!\pi_I)}{(\pi_I\!-\!X_I)^2}
\Bigg\}
.
\end{eqnarray}
This expression simplifies in the isospin symmetric
limit to
\begin{eqnarray}
\lefteqn{\delta B^{\rm disc,SU(2)}_{j}(m_u\!=\!m_d) =}
\nonumber\\
&& \pm \frac{1}{2(4\pi f_2)^2}
\left[\ell(X_I)+ (D_I-X_I)\tilde\ell(X_I)\right]
.
\label{eq:BdiscSU2iso}
\end{eqnarray}

There are two striking features of these results.
First, the chiral logarithms for $B_2$ and $B_3$
have exactly the same form but opposite sign to those
for $B_4$ and $B_5$. This is not true for the SU(3) result.
Second, the chiral logarithms for $B_2$ and $B_3$ 
are identical to those found for
$B_K$ in Ref.~\cite{Bae:2010ki}.\footnote{%
This holds for the full $1+1+1$ theory, although the results of
Ref.~\cite{Bae:2010ki} enable a check only for the $2+1$ flavor theory.}
These features will be explained by the following analysis.

\bigskip
We now turn to the arguments which justify the prescription
used to obtain the above results.
The assumption made in Ref.~\cite{Bae:2010ki} is that
the NLO SU(2) SChPT results can be obtained by
working in SU(3) SChPT to all orders in $m_y$ and $m_s$
while working only to NLO in the small quantities $m_x$, $m_u$, $m_d$, $a^2$,
$a_\alpha^2$  and $\alpha^2$.
This assumes that there are no non-perturbative contributions in $m_y$
and $m_s$.
In this approach,
any number of loops of particles containing strange quarks
are allowed, but only a single loop containing light particles.
One then imagines taking the SU(2) limit, so that heavy particle
loops lead only to contact terms, which are analytic in
the light quark masses.

It is important to note that, while loops of heavy particles 
are not suppressed (since they give contributions like
$(m_K/\Lambda_\chi)^2 \ln(m_K/\mu) \sim 1$), they are also
not enhanced in the SU(2) power counting. Thus a contribution
to the NLO SU(3) SChPT result whose suppression is 
by one of the small quantities in SU(2) SChPT
remains at most of NLO in SU(2) power counting
with the addition of any number of heavy loops.
This implies that contributions of NLO in SU(2) power counting
can arise either from a light particle loop
or from an operator bringing in an explicit factor of
$m_x$, $m_u$, $m_d$, $a^2$, $a_\alpha^2$ or $\alpha^2$ 
(in both cases, with any number of heavy loops),
but not from both.
For the analytic contributions to $B_j$, this leads immediately
to the result of Eq.~(\ref{eq:deltaBanalSU(2)}), with
$d_{j1}-d_{j5}$ being arbitrary analytic functions of $m_y$ and $m_s$.

A further observation is that, at NLO in SU(2) power counting, 
where heavy loops are collapsed to a point,
all one-loop diagrams are tadpoles,
i.e. with no vertices on the light particle propagator.
They have the generic form of Figs.~\ref{fig:BKnum}(d) or (e) with
a pion in the loop.
Diagrams of the type of Fig.~\ref{fig:BKnum}(f)
collapse to an analytic contribution since there is necessarily
a kaon in the loop.
This observation implies that, since the loop itself
gives a contribution of NLO in SU(2) power counting,
the vertex to which it attaches must be LO in this power counting.
Since at LO there are no factors of $a^2$, $a_\alpha^2$ or $\alpha^2$
in the operators, 
the vertex must be the same as that obtained 
by taking the SU(2) limit of the continuum result obtained
to all orders in SU(3) ChPT.
By assumption, however, this limit gives the same vertices as
directly working with continuum SU(2) ChPT.
Thus we conclude that the non-analytic NLO contribution
can be obtained by calculating the tadpole vertices using
continuum SU(2) ChPT and inserting the propagators from
the SU(2) limit of SChPT.

The final step of the argument is to note that
in SU(2) ChPT 
{\em the relative contribution of the 
chiral logarithms is independent of the masses $m_y$ and $m_s$}.
This point, derived in Ref.~\cite{rbc-uk-08}, will be explained
briefly below.
This relative contribution is therefore the same as in
a theory with 
$m_{x}\sim m_u\sim m_d \ll m_y \sim m_s \ll \Lambda_{\rm QCD}$.
In such a theory it is justified to obtain the NLO SU(2) SChPT
result by calculating at NLO using SU(3) SChPT and taking
the SU(2) limit, which is exactly the procedure used above
except that the LECs are independent of $m_y$ and $m_s$.
We therefore conclude that the form of the SU(2) SChPT
chiral logarithms obtained above is correct.
We stress again that the LEC $f_2$ does depend on $m_y$ and
$m_s$; what is independent of these masses is the
remainder of the expressions in Eqs.~(\ref{eq:BconnSU2}-\ref{eq:BdiscSU2iso}).

\bigskip
In order to check the final step of this argument,
we have calculated the chiral logarithms in the hybrid
theory described above, in which the vertices 
are from continuum SU(2) ChPT
while the propagators are from SChPT in the SU(2) limit.
This calculation is also useful because it shows why the
results for $B_{2,3}$ ($B_{4,5}$) have exactly the same (opposite)
form to those for $B_K$.

The set-up for continuum SU(2) ChPT in our case requires 
a generalization of the method of Ref.~\cite{rbc-uk-08},
because our continuum theory is the enlarged, partially quenched one
described in Sec.~\ref{sec:Olat_cont}.
In particular, we must account for the extra taste degree of freedom,
the two types of valence $S$ and $D$ quarks,
and the fact that our operators have a different form
[Eqs.~(\ref{eq:O2cont'}-\ref{eq:O5cont'}) compared to
 Eqs.~(\ref{eq:O2cont}-\ref{eq:O5cont})].
These features imply that the kaon fields can be collected
into a rectangular matrix, $K_{ab}\sim \ell_a\bar s_b $,
where the first index runs over
the flavors and tastes of light quarks and ghosts, 
while the second runs over the two types of
strange quark and their tastes.
Thus $a$ takes 24 values  (valence down quarks of both types, corresponding
ghosts, and up and down sea quarks, all with four tastes)
while $b$ takes 8.
The approximate chiral symmetry is now $SU(16|8)_L\times SU(16|8)_R$.
We only know how $K$ transforms under the vector subgroup,
and also under the SU(8) group rotating between
the 8 different strange quarks:
\begin{equation}
K \longrightarrow U K V^\dagger,
\qquad 
U \in SU(16|8)_V, \ V\in SU(8).
\end{equation}
The construction of the SU(2) chiral Lagrangian for pions
follows exactly the same steps as described in
Sec.~\ref{subsec:SChPT}, aside from the absence of $a^2$ terms.
Thus $\Sigma$ is now a $24\times 24$ graded matrix,
transforming as in Eq.~(\ref{eq:Sigmatransform}).
As usual, we couple $K$ into the chiral theory by
introducing
$u = \sqrt \Sigma = \exp[i\Phi/(2f_2)]$, which transforms as
\begin{equation}
u \to L u U^\dagger = U u R^\dagger
.
\end{equation}
Thus the combinations $u K$ and $u^\dagger K$ transform simply
under the full chiral and $SU(8)$ groups
\begin{equation}
(uK) \to L (uK) V^\dagger,\quad
(u^\dagger K) \to R (u^\dagger K) V^\dagger
.
\end{equation}
The leading order kaon Lagrangian is then
\begin{equation}
{\cal L}_{\chi,K} = {\rm tr}\left(
D_\mu K^\dagger D_\mu K + m_K^2 K^\dagger K\right)
,
\end{equation}
where the trace is over the SU(8) indices (which are not graded).
The covariant derivative involves the $u$ fields and is
defined, e.g., in Ref.~\cite{rbc-uk-08}.

To map operators into this SU(2) chiral theory we use a variant
of the spurion method described earlier.
For example, we write one of the component operators of 
${\cal O}_2^{{\rm Cont}'}$ as
\begin{equation}
{\cal O}_{S+P} = \bar S(1\otimes F_{1L})Q_L \bar S(1\otimes F_{2L})Q_L
+ {\rm p.c.}
,
\end{equation}
to be compared with Eq.~(\ref{eq:QformS+P}).
Note the absence of a chirality subscript on the heavy $\bar S$ field.
For economy of notation, we reuse the symbols $F_{1L}$, $Q_L$ etc.,
although they have different meanings in the SU(2) theory.
Here $\bar S$ is a row vector containing the 8 strange quark fields,
while $Q_L$ is a column vector containing the 24 light left-handed
fields and $F_{1L}$ and $F_{2L}$ are $8\times 24$ matrices.
To obtain the desired operator the spurions have to be chosen as
in Eq.~(\ref{eq:setspurions}), where now the Kronecker $\delta$ is
a rectangular matrix.
Under chiral transformations, the light quark fields transform
as in Eq.~(\ref{eq:Qtransform}), while under SU(8) transformations
$\bar S \to \bar S V^\dagger$. Thus the spurions here must
transform as
\begin{equation}
F_{kL} \to V F_{kL} L^\dagger\ \ {\rm and}\ \
F_{kR} \to V F_{kR} R^\dagger
\end{equation}
for $k=1,2$.
The mapping into the chiral theory follows the
same logic as for the SU(3) theory. The simplest operators
that are allowed are
\begin{eqnarray}
{\cal O}_a^{\chi, SU(2)}
&=& 
{\rm tr}(F_{1L} u K) {\rm tr}(F_{2L} u K)
\nonumber\\
&&\quad + {\rm tr}(F_{1R} u^\dagger K) {\rm tr}(F_{2R} u^\dagger K)
\label{eq:OchiaSU2}
\end{eqnarray}
and
\begin{equation}
{\cal O}_b^{\chi, SU(2)}
=
{\rm tr}(F_{1L} u K F_{2L} u K)
+ {\rm tr}(F_{1R} u^\dagger K F_{2R} u^\dagger K)
.
\label{eq:OchibSU2}
\end{equation}
By the same reasoning as before, ${\cal O}_T^{{\rm Cont}'}$
also maps into the same two operators.
Furthermore, calculating the LO matrix elements
between kaon states and enforcing 
(\ref{eq:tasteI}) and (\ref{eq:FierzO}), one finds that the
full operator ${\cal O}_{2}^{{\rm Cont}'}$ maps into
the sum of the two chiral operators.
There is thus only a single overall LEC.
The same form holds for ${\cal O}_3^{{\rm Cont}'}$.

In SU(2) ChPT, operators with additional covariant derivatives
acting on the kaon fields are not suppressed.
As explained in Ref.~\cite{rbc-uk-08}, however,
by using the equations of motion one can reduce these
operators down to those without derivatives, up to
contributions of higher order in the SU(2) power counting.
This is the result that shows how arbitrary powers of
$m_K^2$ can appear without impacting the coupling to pions.

We now sketch how the mapping changes for ${\cal O}_{4,5}^{{\rm Cont}'}$.
Here the quark-level operators, shown in Eqs.~(\ref{eq:QformS-P})
and (\ref{eq:OV-A}),
contain both a $Q_L$ and a $Q_R$ field, in contrast to the $LL$ or
$RR$ structure that appears for ${\cal O}_{2,3}^{{\rm Cont}'}$.
The net result is that one must build the chiral operators
out of $F_{1L}$ and $F_{2R}$ (or their parity conjugates).
Note that, since $S_L$ and $S_R$ transform in the same way,
we do not need to introduce new spurions $\tilde F_{kL}$ and $\tilde F_{kR}$.
We then find that both ${\cal O}_{S-P}$ and ${\cal O}_{V-A}$ are mapped
into a linear combination of
\begin{equation}
{\cal O}_c^{\chi, SU(2)}
= 
{\rm tr}(F_{1L} u K) {\rm tr}(F_{2R} u^\dagger K)
+ {\rm p.c.}
\label{eq:OchicSU2}
\end{equation}
and
\begin{equation}
{\cal O}_d^{\chi, SU(2)}
=
{\rm tr}(F_{1L} u K F_{2R} u^\dagger K)
+ {\rm p.c.}
\label{eq:OchidSU2}
\end{equation}
Enforcing (\ref{eq:tasteI}) and (\ref{eq:FierzO}) we again find that
${\cal O}_{4,5}^{{\rm Cont}'}$ both map into the
sum of these two chiral operators.

The mapping of the bilinears in the denominator of the $B_j$
is simpler and leads to [cf. Eq.~(\ref{eq:Ochibil})]
\begin{equation}
{\rm tr}(F_{kL} u K) + {\rm p.c.}
\end{equation}

Now we observe that the form of the SU(2) ChPT operator
for ${\cal O}_{2,3}^{{\rm Cont}'}$,
i.e. the sum ${\cal O}_c^{\chi, SU(2)}+{\cal O}_d^{\chi,SU(2)}$,
is {\em identical} to the operator which represents the
operator appearing in the numerator of $B_K$,
and whose explicit form is given in Ref.~\cite{Bae:2010ki}.
This is because all three operators have a $LL+RR$ structure in terms
of the light quark fields. Thus the one-loop corrections to
the numerator of $B_{2,3}$ are identical to those for the
numerator of $B_K$. As for the denominators, they have
a somewhat different form (the denominator for $B_K$ is mapped
into SU(2) ChPT in Ref.~\cite{rbc-uk-08}), but at NLO,
and when expressed in terms of $\Phi$ and $K$ fields, they
are proportional. This implies that the chiral logarithms
in both numerator and denominator will be the same
for $B_2$, $B_3$ and $B_K$.
This holds also in the hybrid calculation using
SChPT propagators, since the vertices and the
propagators are the same in all cases.
We thus can take the SU(2) SChPT result from Ref.~\cite{Bae:2010ki}
and indeed find, as observed above, that it agrees
with the sum of Eqs.~(\ref{eq:BconnSU2})
and (\ref{eq:BdiscSU2iso}).
This checks our argument based on taking
the SU(2) limit of the SU(3) SChPT expression.

Finally, we can now understand why the chiral
logarithms in $B_{4,5}$ are, in the SU(2) limit,
exactly opposite to those for $B_{2,3}$ and $B_K$.
The first step is to notice
that the $LR+RL$ structure of the quark-level operators
in $B_{4,5}$ maps into operators at the chiral level with
one $u$ and one $u^\dagger$.
This is in contrast to the two $u$'s or two $u^\dagger$'s
for $B_{2,3}$. 
Next we note that there is a significant cancellation 
of chiral logarithms between the numerators and denominators of the $B_j$.
In particular, as observed for $B_K$ in Ref.~\cite{rbc-uk-08},
the denominator cancels the contributions from the numerator
in which one of the $u/u^\dagger$'s
is expanded out to ${\cal O}(\Phi^2)$, while the other is unity.
(This type of contribution is only non-vanishing for the two trace
chiral operators in the numerator.)
Thus the only contributions surviving the cancellation
come from terms in which both $u$/$u^\dagger$'s are
expanded to linear order in $\Phi$. Recalling
that $u=\exp[i\Phi/(2f)]$, we see that the expansion
of the $u\times u^\dagger$ operators will lead to the opposite
sign to that from $u\times u$ or $u^\dagger\times u^\dagger$.
This implies that the chiral logarithms are opposite
for $B_{2,3}$ and $B_{4,5}$.

\subsection{Continuum PQ results}

As noted in the introduction, the continuum PQChPT result
is not available in the literature, either for SU(3)
or SU(2) ChPT. We can obtain these
results by taking the continuum limit of our general formulae.
In this limit, all taste breakings vanish,
so we can make the substitutions
\begin{equation}
K_B\to K,\quad X_B\to X,\quad Y_B\to Y, \ {\rm etc.},
\end{equation}
as well as setting $a$ and $\alpha$ to zero.

For the SU(3) case, we find
\begin{align}
\delta B^{\rm anal}_j &\to c_{j1} (m_x+m_y)
+ c_{j2} (m_u+m_d+m_s) 
\\
\delta B_j^{\rm conn} &\to
-\frac{
2\ell(K) \!-\! 2 K \tilde\ell(K)
\!\pm\! \ell(X) \!\pm\! \ell(Y) }{(4\pi f)^2}.
\end{align}
These results hold for the general $1+1+1$ flavor
theory. We stress that in the result for
$\delta B_j^{\rm conn}$, $K$ is the squared mass
of the partially quenched kaon (composed of
a quark with mass $m_x$ and an antiquark of mass $m_y$)
as opposed to the physical kaon composed of sea quarks.

For $\delta B_j^{\rm disc}$, the result in the
$2+1$ flavor theory is identical to that in SChPT,
given in Eqs.~(\ref{eq:Bdisc} - \ref{eq:IY}), except
that the subscript $I$ is dropped.
We do not quote the $1+1+1$ flavor result explicitly,
as it is lengthy, but it can be obtained straightforwardly
from Eq.~(\ref{eq:Bdiscsimple}) using
the propagators of Eq.~(\ref{eq:DI}) with the
subscript $I$ dropped.

\bigskip
The result for partially quenched SU(2) continuum
ChPT in the $1+1+1$ flavor theory is
\begin{eqnarray}
B_j &=& B_{j}^{\rm LO,SU(2)}\Bigg[1 +   d_{j1} m_x + d_{j2} (m_u+m_d)
\nonumber\\
&&\pm \frac{1}{2(4\pi f_2)^2}
\Bigg\{
\tilde\ell(X)\frac{(U\!-\!X)(D\!-\!X)}{\pi\!-\!X}
\nonumber\\
&&-
\ell(X)
\frac{(U\!-\!X)(D\!-\!X)}{(\pi\!-\!X)^2}
\nonumber\\
&&+
\ell(\pi)
\frac{(U\!-\!\pi)(D\!-\!\pi)}{(\pi\!-\!X)^2}
\Bigg\}\Bigg]
.
\end{eqnarray}
This reduces in the isospin limit ($m_u=m_d\equiv m_\ell$) to
\begin{eqnarray}
B_j &=& B_{j}^{\rm LO,SU(2)}\Bigg[1 +   d_{j1} m_x + d_{j2} 2 m_\ell
\nonumber\\
&& \pm \frac{1}{2(4\pi f_2)^2}
\left\{ 
\tilde\ell(X)(\pi\!-\!X) - \ell(X) \right\}
.
\end{eqnarray}

\section{Conclusions}
\label{sec:conc}

We have presented the next-to-leading order results in staggered chiral
perturbation theory for $B$-parameters of the kaon mixing operators that 
generically arise in models of new physics.
We have done so for both SU(3) and SU(2) chiral perturbation theory.
These results can be used to extrapolate lattice data obtained using
staggered fermions to the physical light and strange quark masses.
As a side product, we also provide partially quenched
results for both SU(3) and SU(2) ChPT in the continuum.

We find that the results are much simpler in
SU(3) SChPT than for $B_K$. Terms induced by discretization
and matching errors in the lattice operators enter only
through analytic terms rather than through chiral logarithms.
We also find that the use of a mixed action does not change
the form of the NLO results.
For SU(2) SChPT the results are of comparable simplicity to those
for $B_K$.
Indeed, the chiral logarithms for $B_2$ and $B_3$ are
identical to those for $B_K$, while those for $B_4$ and $B_5$
are opposite.
In both SU(3) and SU(2) SChPT, if one works at fixed lattice
spacing, the NLO expressions have the same number of
unknown constants as those in the continuum,
as long as one first determines the masses of the valence pions
and kaons of all tastes.

It was pointed out in Ref.~\cite{Becirevic:2004qd} that certain combinations
of $B$-parameters have vanishing or small chiral logarithms.
The former combinations, dubbed ``golden'' in Ref.~\cite{Becirevic:2004qd}, 
remain golden in SChPT. 
The two examples built from $B$-parameters alone
are the ratios $B_2/B_3$ and $B_4/B_5$. 
The ``silver'' combinations are
$(\textrm{one of}\ B_{2,3})\times (\textrm{one of}\ B_{4,5})$,
for which the SU(3) chiral logarithms largely cancel.
These turn out to be golden in SU(2) SChPT.
It may be useful to use these combinations to improve
the chiral extrapolations.

\section{Acknowledgments}
The research of W.~Lee is supported by the Creative Research
Initiatives program (3348-20090015) of the NRF grant funded by the
Korean government (MEST).
The work of S.~Sharpe is supported in part by the US DOE grant
no.~DE-FG02-96ER40956. J.~Bailey, H.-J.~Kim and W.~Lee thank the
University of Washington for hospitality while this work was begun.
%
\bibliographystyle{apsrev} 
\bibliography{ref} 

\end{document}